\pdfoutput=1
\newcommand*{\ATLASLATEXPATH}{latex/}
\documentclass[cernpreprint,texlive=2016,txfonts,UKenglish]{\ATLASLATEXPATH atlasdoc}
\usepackage[subfigure]{\ATLASLATEXPATH atlaspackage}
\usepackage{\ATLASLATEXPATH atlasbiblatex}

\usepackage{\ATLASLATEXPATH atlascontribute}

\usepackage{\ATLASLATEXPATH atlasphysics}

\graphicspath{{logos/}{figures/}}

\usepackage{ATLASpaper-defs}
\usepackage{multirow}
\usepackage{epstopdf}

\graphicspath{{logos/}}

\AtlasTitle{Measurement of inclusive jet and dijet cross-sections in proton\texttt{--}proton collisions at $\sqrt{s}=13$\,\TeV{} with the ATLAS detector}

\PreprintIdNumber{CERN-EP-2017-157}

\AtlasJournalRef{JHEP 05 (2018) 195}

\AtlasDOI{https://doi.org/10.1007/JHEP05(2018)195}

\AtlasAbstract{%
	Inclusive jet and dijet cross-sections are measured in proton\texttt{--}proton collisions at a centre-of-mass energy of 13\,\TeV. The measurement uses a dataset with an integrated luminosity of 3.2\,fb$^{-1}$ recorded in 2015 with the ATLAS detector at the Large Hadron Collider. Jets are identified using the \AKT algorithm with a radius parameter value of $R=0.4$. The inclusive jet cross-sections are measured double-differentially as a function of the jet transverse momentum, covering the range from \ptMin{} to \ptMax{}, and the absolute jet rapidity up to $|y|=3$. The double-differential dijet production cross-sections are presented as a function of the dijet mass, covering the range from \mjjMin{} to \mjjMax{}, and the half absolute rapidity separation between the two leading jets within $|y|<3$, $\ystar$, up to $\ystar=3$. 
Next-to-leading-order, and next-to-next-to-leading-order for the inclusive jet measurement, perturbative QCD calculations corrected for non-perturbative and electroweak effects are compared to the measured cross-sections.
}

\hypersetup{pdftitle={ATLAS document},pdfauthor={The ATLAS Collaboration}}

\begin{document}

\maketitle

\tableofcontents

\clearpage

\section{Introduction}
\label{sec:intro}
Precise measurements of jet cross-sections are crucial in understanding physics at hadron colliders. In quantum chromodynamics (QCD), jets are interpreted as resulting from the fragmentation of quarks and gluons produced in a short-distance scattering process. Jet cross-sections provide valuable information about the strong coupling constant, $\alpha_\text{s}$, and the structure of the proton. Also, inclusive jet and dijet events represent a background to many other processes at hadron colliders. The predictive power of fixed-order QCD calculations is therefore relevant in many searches for new physics.
 
Inclusive jet production cross-sections have been measured in collisions of hadrons at the $\rm{Sp\bar{p}S}$ and Tevatron colliders at various centre-of-mass energies.
The latest and most precise results at $\sqrt{s} = 1.96$\,\TeV{} are detailed in Refs.~\cite{Aaltonen:2008eq,Abazov2012}.
At the Large Hadron Collider (LHC)~\cite{Evans:2008zzb} at CERN, the ALICE, ATLAS and CMS collaborations have measured inclusive jet cross-sections in proton\texttt{--}proton collisions at centre-of-mass energies of $\sqrt{s} =  2.76$\,\TeV~\cite{ALICEIJXS2013,STDM-2012-08,CMS-SMP-14-017} and $\sqrt{s} = 7$\,\TeV~\cite{STDM-2010-03,CMS-QCD-11-004,CMS-SMP-13-002,CMS-QCD-10-011}.
Recently, the ATLAS and CMS collaborations have measured the inclusive jet cross-sections at $\sqrt{s} = 8$\,\TeV~\cite{STDM-2015-01,CMS-SMP-14-001}, and the CMS Collaboration also at $\sqrt{s} = 13$\,\TeV~\cite{CMS-SMP-15-007}. 
Dijet production at the LHC has been measured by the ATLAS and CMS collaborations at a centre-of-mass energy of $\sqrt{s} = 7$\,\TeV~\cite{STDM-2010-03,CMS-QCD-10-025}, and also by the CMS Collaboration at $\sqrt{s} = 8$\,\TeV~\cite{CMS-SMP-16-011}.

This paper presents measurements of the inclusive jet and dijet cross-sections in proton\texttt{--}proton collisions at $\sqrt{s}=13$\,\TeV{} centre-of-mass energy by the ATLAS Collaboration at the LHC, using data collected in 2015 and corresponding to an integrated luminosity of $3.2$\,fb$^{-1}$. The inclusive jet cross-sections are measured double-differentially as a function of the jet transverse momentum, \pt{}, and absolute jet rapidity, $|y|$.\footnote{The rapidity is defined as $y=0.5\times \ln[(E+p_z)/(E-p_z)]$, where $E$ denotes the energy and $p_z$ is the component of the momentum along the beam direction.} In addition, the double-differential dijet production cross-sections are presented as a function of the invariant mass of the dijet system, $\twomass{j}{j}$, and as a function of half the absolute rapidity separation between the two highest-\pt{} jets satisfying $|y|<3$, denoted $\ystar$.\footnote{The variable $y^{*}$ is defined as $\left|y_1-y_2\right|/2$, where the subscripts 1,2 label the highest and second highest-\pt{} jet in the event satisfying $|y|<3.0$, respectively. This quantity is invariant under a Lorentz boost along the $z$-direction and is equal to the absolute rapidity of each jet in the dijet rest frame.} Jets are reconstructed using the \antikt jet clustering algorithm~\cite{Cacciari:2008gp} with a radius parameter value of $R = 0.4$. The measurements cover the kinematic region of $\ptMin<\pt<\ptMax$ and $|y|<3$ for the inclusive jet cross-section, and of $\mjjMin<\twomass{j}{j}<\mjjMax$ and $\ystar<3$ for the dijet cross-section.

Next-to-leading-order (NLO) perturbative QCD (pQCD) predictions calculated using several parton distribution function (PDF) sets,
corrected for electroweak and non-perturbative effects, are quantitatively compared to the measurement results. In addition, the inclusive jet cross-sections are compared to the recently published complete next-to-next-to-leading-order (NNLO) pQCD calculation~\cite{NNLO,Currie:2017ctp}.

\newpage

\section{ATLAS detector}
\label{sec:atlas}
The ATLAS experiment~\cite{PERF-2007-01,ATLAS-TDR-19} at the LHC is a multi-purpose particle detector with a forward-backward symmetric cylindrical geometry and a near 4$\pi$ coverage in solid angle.\footnote{ATLAS uses a right-handed coordinate system with its origin at the nominal interaction point (IP) in the centre of the detector and the $z$-axis along the beam pipe. The $x$-axis points from the IP to the centre of the LHC ring, and the $y$-axis points upward. Cylindrical coordinates $(r,\phi)$ are used in the transverse plane, $\phi$ being the azimuthal angle around the $z$-axis. The pseudorapidity is defined in terms of the polar angle $\theta$ as $\eta=-\ln\tan(\theta/2)$.} 
It consists of an inner tracking detector, electromagnetic and hadron calorimeters, and a muon spectrometer. The inner tracking detector covers the pseudorapidity range $|\eta|<2.5$ and is used to reconstruct tracks and vertices. It consists of silicon pixel, silicon microstrip, and transition radiation tracking detectors, surrounded by a thin superconducting solenoid providing a 2\,T axial magnetic field.
Lead/liquid-argon (LAr) sampling calorimeters provide electromagnetic (EM) energy measurements with high granularity. They consist of a barrel ($|\eta| < 1.475$) and two endcap ($1.375 \le |\eta| < 3.2$) regions. The hadron calorimeters are divided into five distinct regions: a barrel region ($|\eta| < 0.8$), two extended barrel regions ($0.8 \le |\eta| < 1.7$) and two endcap regions ($1.5 \le |\eta| < 3.2$). The barrel and extended barrel regions are instrumented with steel/scintillator tile calorimeters. The endcap regions are instrumented with LAr calorimeters for both the EM and hadronic energy measurements. The ATLAS calorimeters have very high lateral granularity and several samplings in depth over $|\eta| < 3.2$. The muon spectrometer surrounds the calorimeters and features three large air-core toroid superconducting magnets with eight coils each. The field integral of the toroids ranges between 2.0 and 6.0\,Tm across most of the detector. It includes a system of precision tracking chambers for track measurement in the principal bending direction and fast detectors for triggering and measurement of the muon coordinate in the direction orthogonal to that determined by the precision-tracking chambers. A two-level trigger system is used to select events. The first-level trigger is implemented in hardware and uses a subset of the detector information. This is followed by the high-level trigger system~\cite{TRIG-2016-01}, which is software-based and can run the offline reconstruction and calibration software, further reducing the event rate to an average of 1\,kHz.

\section{Cross-section definitions}
\label{sec:definitions}
The jet cross-sections are determined for so-called particle jets. These jets are built at the event generator level from stable particles, i.e.\ those fulfilling $c\tau > 10$\,mm, where $\tau$ is the proper lifetime. This definition includes muons and neutrinos. Jets are identified using the \AKT jet algorithm~\cite{Cacciari:2008gp} as implemented in the \fastjet~\cite{FastJet} package with radius parameter $R=0.4$. The use of the \AKT algorithm is well motivated since it is infrared- and collinear-safe, and produces geometrically well-defined (``cone-like'') jets.

Inclusive jet double-differential cross-sections are measured as a function of jet $\pt$ in six equal-size bins of the absolute jet rapidity, $|y|$. Only jets in the kinematic range $\pt > 100\,\gev$ and $|y| < 3.0$ are considered, to ensure that the jet energy scale is well understood, as described in Section~\ref{sec:jesjer}. The inclusive jet production cross-section can be expressed as a ratio of the number of jets in data after correcting for detector effects, $N_\text{jets}$, to the integrated luminosity of the data, $\mathcal{L}$, in a given interval of momentum and rapidity, $\Delta\pt$ and $\Delta y$ respectively:
\begin{equation}\nonumber
	\frac{{\mathrm{d}}^2\sigma}{\mathrm{d}\pt \mathrm{d}y} = \frac{N_\text{jets}}{\mathcal{L}\Delta\pt\Delta y}.
\end{equation}

The dijet double-differential cross-section is measured as a function of the invariant mass of the dijet system, $\twomass{j}{j}$, in six equal-size bins of \ystar, for events with at least two jets with $\pt>75$\,\GeV{} and $|y|<3.0$. In addition, the scalar sum of the $\pt$ of the first and second leading jets, $H_{\mathrm{T},2} = {\pt}_{1} + {\pt}_{2}$, is required to be above $200$\,\GeV{}. This requirement avoids instabilities in the NLO cross-section calculations due to the symmetric \pt{} requirement applied to the leading and sub-leading jets~\cite{Frixione:1997ks,POWHEGHT2}.

The dijet production cross-section is calculated as the ratio of the number of dijet events after correcting for detector effects, $N_\text{dijet}$, to the integrated luminosity of the data, in a given interval of the invariant mass and \ystar{}, $\Delta\twomass{j}{j}$ and $\Delta\ystar$ respectively:
\begin{equation}\nonumber
	\frac{{\mathrm{d}}^2\sigma}{\mathrm{d}\twomass{j}{j} \mathrm{d}\ystar} = \frac{N_\text{dijet}}{\mathcal{L}\Delta\twomass{j}{j}\Delta\ystar}.
\end{equation}

The $H_{\mathrm{T},2}$ and $\pT$ selections on the two leading jets determine the lower edge of the mass range in each \ystar{} bin.
The \pt{} ($\twomass{j}{j}$) binning is chosen according to the detector \pt{} ($\twomass{j}{j}$) resolution, such that the bin width is approximately twice the \pt{} ($\twomass{j}{j}$) resolution, with the exception of the highest \pt{} ($\twomass{j}{j}$) bins in each rapidity (\ystar) range where the bin width is enlarged to avoid large statistical fluctuations and non-Gaussian statistical uncertainties due to a low number of entries per bin; as predicted by a MC simulation (see Section~\ref{sec:dataANDmc}).

\section{Dataset and Monte Carlo simulations}
\label{sec:dataANDmc}
The measurements use proton\texttt{--}proton collision data at a centre-of-mass energy of $\sqrt{s} = 13$\,\TeV{} collected by the ATLAS detector during the 2015 data-taking period of the LHC. The LHC beams were operated with proton bunches organised in ``bunch trains'', with a bunch spacing of 25\,ns. The integrated collected luminosity is $3.2\,\text{fb}^{-1}$ with an uncertainty of 2.1\%. The uncertainty in the luminosity is derived following a methodology similar to that detailed in Ref.~\cite{DAPR-2013-01}, from a calibration of the luminosity scale using $x$\texttt{--}$y$ beam-separation scans performed in August 2015.

Simulated jet events were produced using three different Monte Carlo event generators for comparisons to data and to derive corrections. The \pythiaeight{} program (version 8.186~\cite{Sjostrand:2007gs}) was used for the baseline comparisons, the deconvolution of detector effects and the propagation of systematic uncertainties. It uses LO pQCD matrix elements for $2 \rightarrow 2$ processes, along with a leading-logarithmic (LL) \pT-ordered parton shower~\cite{Sjostrand2005} including photon radiation, underlying event\footnote{The term underlying event is used to mean particles produced in the same proton\texttt{--}proton collision, but not originating from the primary hard partonic scatter or its products.} simulation with multiple parton interactions~\cite{1126-6708-2004-03-053}, and hadronisation with the Lund string model~\cite{ANDERSSON198331}. The samples were created using a set of tuned parameters called the A14 tune~\cite{ATL-PHYS-PUB-2014-021} and the NNPDF2.3LO~\cite{NNPDF23} LO PDF set.  The EvtGen 1.2.0 program~\cite{Lange2001152} was used to model bottom and charm hadron decays.
NLO samples of simulated events were produced using \powheg~\cite{Nason:2004rx,POWHEGHT2} and showered with \pythiaeight{} for systematic studies as discussed in Section~\ref{sec:jesUncert}, and for optimising bin widths.
The A14 tune and the CT10~\cite{Lai:2010vv} PDF set were used. 
For the evaluation of non-perturbative effects, the \pythiaeight{} and \herwigpp{}~\cite{Bahr:2008pv} (v2.7.1~\cite{Bellm:2013lba}) event generators were also employed as described in Section~\ref{sec:npc}.

In all the samples, the effects of multiple proton\texttt{--}proton interactions in the same and neighbouring bunch crossings (pile-up) were included by overlaying inelastic minimum-bias events generated with \pythiaeight. The generated events were further weighted to reproduce the observed distribution of the average number of collisions per bunch crossing in data. The stable particles from the generated events were passed through the ATLAS detector simulation~\cite{SOFT-2010-01} based on \geant~\cite{Agostinelli:2002hh} and  were  reconstructed  with the  same  version  of  the  ATLAS software as was used to process the data.

\section{Event and jet selection}
\label{sec:selection}
A suite of single-jet triggers~\cite{TRIG-2016-01} with thresholds varying from 55\,\GeV{} to 360\,\GeV{} are used to record events with at least one jet with transverse energy, \ET{}, above the threshold in the region $|\eta| < 3.2$. To keep the trigger rate to an acceptable level, the triggers with lower \ET{} thresholds are prescaled by recording only a predefined fraction of events. The highest-threshold trigger accepts all events passing the threshold. 
The effective luminosities range from $81\,\text{nb}^{-1}$ for $75<\pt<100$\,\GeV, where the trigger prescaling is largest, to $3.2\,\text{fb}^{-1}$ for $\pt>442$\,\GeV, where an unprescaled trigger is used.

A \pt{}-dependent trigger strategy is adopted in order to optimise the statistical power of the measurement. In the inclusive jet measurement, each \pt{} bin requires the trigger with the lowest prescale (i.e. with highest effective luminosity) that is fully efficient in that range.
Due to the high prescale factors, trigger efficiencies are studied offline in data as a function of \pt{} and rapidity by emulating the online trigger decision. The efficiency for jets in a given \pT{} range is obtained as the fraction of those that pass the emulated trigger in an unbiased sample obtained by requiring at least one online jet passing the first-level trigger with $\ET > 15$\,\GeV. The trigger efficiency is always larger than 99.9\% in the \pt{} range where it is considered.

The trigger strategy for the dijet measurement is slightly different to account for different prescale combinations for dijet events in a given ($\twomass{j}{j},\ystar$) bin, which can be accepted by up to two jet triggers depending on the transverse momenta and pseudorapidities of the leading and sub-leading jets. Each pairing of triggers has a unique corresponding luminosity, which is used to calculate the differential cross section for that pairing. The separate cross sections from all pairings are then summed to obtain the final measurement. This strategy is described in detail in Ref.~\cite{Lendermann:2009ah}. It was carefully validated in the previous dijet analysis~\cite{STDM-2010-03} using dedicated simulation samples containing a complete set of prescaled triggers, similar to those used in this measurement.

Events are required to have at least one reconstructed vertex with at least two associated well-reconstructed tracks. The vertex maximising $\sum\pt^2$, where the sum is performed over the associated tracks, is chosen as the primary vertex.
Quality criteria are applied to reject events with jets from beam-induced background due to proton losses upstream of the interaction point, cosmic-ray air showers overlapping with collision events and calorimeter noise from large-scale coherent noise or isolated pathological cells. These jet cleaning criteria are described in Ref.~\cite{ATLAS-CONF-2015-029}.

\section{Jet energy calibration and resolution}
\label{sec:jesjer}
\subsection{Jet reconstruction}

The input objects to the jet algorithm are three-dimensional topological clusters (topoclusters)~\cite{PERF-2011-03,ATL-LARG-PUB-2008-002} built from the energy deposits in calorimeter cells.\ A calibration is applied to the clusters to give the correct response for the energy deposited in electromagnetic showers, while it does not correct for the lower response to hadronic showers. The four-momentum of a jet is defined as the sum of the four-momenta of its clusters in the calorimeter, treating each cluster as a four-momentum with zero mass.

\subsection{Jet energy calibration}

Jets in data and simulation are calibrated following the procedure described in Ref.~\cite{PERF-2016-04}. The four-momenta of the jets are recalculated to originate from the hard-scatter vertex rather than from the centre of the detector. The jet energy is corrected for the effect of pile-up in both the collision data and simulated events using the methods described in Ref.~\cite{PERF-2014-03}. In addition, a jet energy- and $\eta$-dependent correction is applied to reconstructed jets in data and Monte Carlo (MC) simulation.\ It is derived from MC simulation and is designed to lead to agreement in energy and direction between reconstructed jets and particle jets on average. Further corrections are applied sequentially (Global Sequential Calibration~\cite{ATLAS-CONF-2015-002}) using five jet substructure variables to reduce effects from fluctuations in the flavour composition of particles forming the jets and fluctuations in the hadronic shower caused by interactions of the hadrons with dead material in the calorimeter. Differences in energy response between data and simulation are evaluated using in situ techniques, where the \pt{} of the jet to be calibrated is balanced against well-measured objects. The full jet energy scale (JES) calibration procedure and its associated systematic uncertainties are described in more detail in the following.

\textbf{Pile-up correction:} Jets are corrected for the contributions from additional proton\texttt{--}proton interactions within the same (in-time) or nearby (out-of-time) bunch crossings~\cite{PERF-2014-03}. First, a correction based on the jet area and the average transverse energy density of the event is derived~\cite{CacciariSalam}. The jet area is a measure of the susceptibility of the jet to pile-up and is determined jet by jet, while the average energy density serves as a measure of the pile-up activity and is calculated event by event with $k_t$-jets with a radius parameter value of $R=0.4$. After this correction, some dependence of the average jet \pt{} on pile-up activity remains. An additional correction is therefore derived by comparing reconstructed calorimeter jets to particle jets in simulated inclusive jet events. The correction is parameterised as a function of the mean number of interactions per bunch crossing, $\mu$, and the number of reconstructed primary vertices in the event, $N_{\mathrm{PV}}$, such that both the out-of-time and in-time effects are taken into account.

The correction for contributions from additional proton\texttt{--}proton interactions can also remove part of the soft-physics contributions to the jet energy, e.g.\ that from the underlying event. This contribution is restored on average by the MC-based jet energy scale correction discussed below.

\textbf{Jet energy scale:} This calibration is derived as a function of the energy and pseudorapidity of the jet using simulated samples of inclusive jet events. The jet energy and pseudorapidity are corrected for instrumental effects (non-compensating calorimeter response, energy losses in dead material and out-of-cone effects) so that they agree on average with the energy and direction of the matching particle jet.

\textbf{Global sequential calibration:} The topology of the energy deposits in the calorimeter and of the tracks associated with the jets is exploited to correct for fluctuations in the jet's particle content~\cite{PERF-2016-04,ATLAS-CONF-2015-002}. The calibration is based on the number of tracks, on the \pT{}-weighted average angular distance between the tracks and the calorimeter jet axis, on the longitudinal extent of the shower in the calorimeter and on the number of track segments in the muon spectrometer associated with the jet. This correction is performed such that the jet energy scale is unaltered on average, but the jet energy resolution is improved and the sensitivity to jet fragmentation effects such as differences between quark- or gluon-induced jets is reduced.

\textbf{In situ techniques:} An in situ calibration is derived to correct for remaining differences between the jet energy response in data and simulation. This correction is calculated using $\gamma+$jet, $Z+$jet, dijet and multijet \pt-balance techniques~\cite{PERF-2016-04,ATLAS-CONF-2015-017,ATLAS-CONF-2015-057}. Up to a jet \pt{} of about 950\,\GeV, the \pt{} balance between a photon or a $Z$ boson and a jet is exploited. The multijet \pt-balance technique calibrates high-\pt{} jets ($300<\pt<2000$\,\GeV) recoiling against a collection of lower-\pt{} jets. Beyond $2000$\,\GeV{} the response is considered constant. All these corrections are derived for the central jets, with $|\eta|<1.2$. The relative response of all detector regions is equalised using a \pt-balance method exploiting dijet events ($\eta$-intercalibration) where the two leading jets are in different $\eta$-regions.

\subsection{Jet energy scale uncertainties}\label{sec:jesUncert}

The jet corrections are combined following the procedure described in Refs.~\cite{PERF-2016-04,ATLAS-CONF-2015-037}. The systematic and statistical uncertainties of each of the above-mentioned calibration steps contribute to the total JES uncertainty as independent systematic components.

Differences between the calorimeter responses to jets initiated by quarks or gluons and a lack of knowledge of the flavour composition of the analysed data lead to additional uncertainties. In order to reduce this contribution, \pythiaeight{} and \powheg{}+\pythiaeight{} Monte Carlo simulations are used to estimate the flavour composition of the sample as a function of \pt{} and rapidity. The result from \pythiaeight{} is taken as the nominal quark/gluon composition, and the difference between the two simulations as an estimate of the composition uncertainty.

A systematic uncertainty is also considered for the muon-segment-based correction, derived as the maximum difference in the jet response between data and MC dijet events as a function of the number of muon segments~\cite{ATLAS-CONF-2015-002}.

An uncertainty in the jet energy scale at high-\pt{}, for jets where in situ methods cannot be used, is derived from single-particle response measurements~\cite{PERF-2015-05}.

Four uncertainties are included to account for potential mismodelling of pile-up in the MC simulation: the number of reconstructed primary vertices, $N_{\mathrm{PV}}$, the average number of interactions per bunch crossing, $\mu$, the energy density in jets and the residual dependence of the jet \pt{} on pile-up. 
The description and evaluation of the pile-up uncertainties are described in detail in Refs.~\cite{PERF-2016-04,PERF-2014-03}.

The measurements presented in this paper use the most detailed description of the systematic uncertainties considered in ATLAS. There are, in total, 76 independent sources of systematic uncertainty treated as being uncorrelated among each other~\cite{PERF-2016-04}. All of these are treated as being fully correlated across \pt{} and $\eta$, with the exception of the statistical uncertainty of the $\eta$-intercalibration which is propagated as being uncorrelated between the 245 different $\eta$ and \pt{} bins in which it was derived~\cite{PERF-2016-04,ATLAS-CONF-2015-017}.
The JES uncertainty is 1\% in the $200-600$\,\GeV{} range of jet \pt, 2\% at $2$\,\TeV, and reaches 3\% above $3$\,\TeV. The uncertainty is fairly constant as a function of $\eta$ and reaches  2.5\% at 80\,\GeV{} for the most forward jets~\cite{PERF-2016-04}.

\subsection{Jet energy resolution and its uncertainties}

The fractional uncertainty in the jet \pt{} resolution (JER) is derived using the data collected during 2012. It is obtained in situ from the standard deviation of the ratio of the \pt{} of a jet to the \pt{} of other well-measured objects (a photon or a $Z$ boson~\cite{ATLAS-CONF-2015-057,ATLAS-CONF-2015-017}) in an event, following techniques similar to those used to determine the JES uncertainty. The \pt{}-balance technique in dijet events ($\eta$-intercalibration)~\cite{ATLAS-CONF-2015-017} allows a measurement of the JER at high jet rapidities and for a wide range of transverse momenta. Noise from the calorimeter electronics and pile-up forms a significant component of the JER at low \pt{}.\ A study in zero-bias data \footnote{The zero-bias sample contains data collected by recording events exactly one accelerator turn after a high \pt{} first-level calorimeter trigger.  These events will thus be contained in a random filled bunch collision with a rate proportional to the instantaneous luminosity \cite{ATLAS-CONF-2015-037}.} allows this contribution to be constrained.
In addition, a MC simulation is used in each in situ JER to correct for fluctuations present at particle level due to the underlying event and out-of-cone contributions from QCD radiation and hadronisation. The results from all these methods are combined in a way similar to that for the JES~\cite{ATLAS-CONF-2015-037}. 

The JER uncertainty has in total 11 components. Eight of these components are obtained by combining the systematic uncertainties associated to the in situ methods. One component is the uncertainty due to the electronic and pile-up noise measurement. Another is the absolute JER difference between data and MC simulation as determined with the in situ methods. Finally, the JER uncertainties are completed with an extra component to account for the differences between the 2012 and 2015 data-taking conditions~\cite{ATL-PHYS-PUB-2015-015}. Each JER systematic component describes an uncertainty that is taken to be fully correlated in jet \pt{} and $\eta$. The 11 JER components are treated as fully uncorrelated with each other.

\subsection{Jet angular resolution and its uncertainties}

The jet angular resolution (JAR) is estimated in MC simulation from the differences in rapidity and azimuthal angle between reconstructed
jets and matching particle jets.  This estimate is validated
by comparing the standard jets built from calorimeter energy deposits to those built from tracks in the inner detector~\cite{PERF-2012-01,PERF-2011-03}. From these studies, the JAR is assigned an uncertainty of 10\% to account for possible differences between data and MC simulation.

\section{Unfolding of detector effects}
\label{sec:unfolding}
The reconstructed jet spectra in data are corrected for detector inefficiencies and resolution effects to obtain inclusive jet and dijet cross-sections that refer to the stable particles entering the detector. The unfolding of the detector resolution in jet \pt{} is based on a modified Bayesian technique, the iterative dynamically stabilised (IDS) method~\cite{Malaescu:2011yg}. This unfolding method uses a transfer matrix constructed using samples of simulated events, which describes the migrations of jets (events) across $\pt$ $(\twomass{j}{j})$ bins between particle level jets and reconstructed level jets. For the inclusive jet measurement, the transfer matrix is filled jet by jet by matching a particle jet with a reconstruction level jet, when both are closer to each other than to any other jet, lie within a radius of $R = 0.3$, have $\pt>75$\,\GeV{} and belong to the same rapidity bin. For the dijet case, the transfer matrix is filled event by event with those events that lie in the same \ystar{} bin and pass the selection requirements at both the reconstruction and the particle levels.

The unfolding technique is performed in three steps, correcting for the matching impurity at the reconstruction level, the smearing of matched jets (events) between \pT{} $(\twomass{j}{j})$ bins, and the matching inefficiency at the particle level,

\begin{equation*} 
	N^\text{part}_i = \sum_k N^\text{reco}_k \cdot \mathcal{P}_k \cdot \mathcal{U}_{ik}\,/\,\mathcal{E}_i\,,
\end{equation*}
where $i$ and $k$ are the \pt ($\twomass{j}{j}$) bin indices of the jets (events) at the particle and reconstruction levels and $N^\text{part}$ and $N^\text{reco}$ are the numbers of particle level and reconstruction level jets (events) in a given bin. The symbols $\mathcal{P}$ and $\mathcal{E}$ denote respectively the matching purity and the matching efficiency. The symbol $\mathcal{U}$ denotes the unfolding matrix, where $\mathcal{U}_{ik}$ describes the probability for a jet (event) at reconstruction level in \pt{} $(\twomass{j}{j})$ bin $k$ to originate from the particle level in \pt{} $(\twomass{j}{j})$ bin~$i$.

For the inclusive jet cross-section measurements, the matching purity, $\mathcal{P}_k$, is defined as the fraction of reconstruction level jets that are matched to a particle level jet for a given \pt{} bin $k$. The matching efficiency, $\mathcal{E}_i$, is defined as the fraction of particle level jets that are matched to a reconstruction level jet for a given \pt{} bin $i$. If matched particle and reconstructed jets are in different rapidity bins then they are reassigned as being unmatched. 
For dijets, the efficiency (purity) is defined as the fraction of events passing the selection cuts at the particle (reconstruction) level for a given \ystar{} bin that also pass the selection cuts and lie in the same \ystar{} bin at the reconstruction (particle) level. In this way the migrations across jet $|y|$ and dijet \ystar{} bins are effectively taken into account by bin-to-bin corrections.
The jet matching efficiency is 98\% (96\%) at $\pt=100$\,\GeV{} for low (high) jet rapidity, and reaches 99.7\% at high \pt. The event dijet efficiency is 97\% (85\%) at $\twomass{j}{j}=300$\,\GeV{} ($\twomass{j}{j}=1700$\,\GeV) for low (high) \ystar, and reaches 99.7\% at the highest dijet mass.
 
The unfolding matrix $\mathcal{U}$ depends on the details of the MC model, given that the transfer matrix is used to build it. This model improves when iterated, 
where the number of iterations is chosen such that the residual bias is within a tolerance of 1\% in the bins with less than 10\% statistical uncertainty. The residual bias is evaluated through a data-driven closure test~\cite{STDM-2011-02,Malaescu:2011yg},
in which the particle level spectrum in the MC simulation is reweighted to improve agreement between data and reweighted MC events in the reconstruction level spectra. The ratio of the spectra unfolded with reweighted and nominal MC simulation provides an estimate of the unfolding bias. 
In these measurements only one iteration is used, achieving an uncertainty bias of the order of a few per mille, except at high \pt{} ($\sim1\,\TeV$) and high rapidity where it increases to 5\%.

\section{Propagation of the uncertainties to the cross-sections}
\label{sec:propunc}
The statistical uncertainties are propagated through the unfolding procedure using an ensemble of 1000 pseudo-experiments.\ Each pseudo-experiment is constructed by reweighting each event in data and simulation according to a Poisson distribution with expectation value equal to one. This procedure preserves the correlations between jets produced in the same event. The unfolding is performed for each pseudo-experiment and a covariance matrix is constructed for the cross-section in each $|y|$ or \ystar{} bin. The total statistical uncertainty is obtained from the covariance matrix, where bin-to-bin correlations are also encoded. The separate contributions from the data and from the MC statistics are obtained from the same procedure by reweighting either the data or the simulated events.

All components of the JES uncertainty (see Section~\ref{sec:jesjer}) are propagated through the unfolding procedure using pseudo-data (MC simulations) to avoid the impact of the larger statistical fluctuations in data. The jet \pt{} in pseudo-data is scaled up and down by one standard deviation of each component. This procedure takes into account the correlations between various phase-space regions. The resulting pseudo-data spectra are unfolded for detector effects using the nominal unfolding matrix. The difference between the nominal unfolded cross-section and the systematically shifted unfolded cross-section is taken as a systematic uncertainty. 
The jet energy scale is the dominant uncertainty for $\pt<2500$\,\GeV{} ($\pt<700$\,\GeV) in the first (last) rapidity bin for the inclusive jet measurement, and for $\twomass{j}{j}<4000$\,\GeV{} in the first \ystar{} bin for the dijet mass measurement. In the complementary regions, including the whole $\twomass{j}{j}$ range for the last \ystar{} bin, the dominant source of uncertainty is the limited size of the sample.

The uncertainty in the JER is the second largest individual source of systematic uncertainty. There are 11 components, some of which can involve a JER degradation in part of $\pt-\eta$ phase-space and a JER improvement in the complementary part, which allows (anti-)correlations to be accounted for.  The effect of each of the components is evaluated by smearing the energy of the reconstructed jets.  The degradation of the JER is achieved by smearing the reconstructed jets in the relevant phase space region in the MC simulation used as pseudo-data. On the other hand, an effective improvement of the JER is achieved by smearing the energy of the jets in the MC simulation used in constructing the transfer matrix. The difference between the modified spectrum unfolded with the systematically varied transfer matrix to the nominal spectrum unfolded with the nominal transfer matrix is taken as a systematic uncertainty.

An uncertainty for the jet cleaning procedure described in Section~\ref{sec:selection} is estimated by measuring in situ the jet selection efficiency.

The uncertainty in the luminosity measurement of 2.1\% is propagated as being correlated across all measurement bins. 

An uncertainty in the beam energy of 0.1\%~\cite{BeamEnergy} is considered when comparing data with the theory prediction at a fixed beam energy. The induced uncertainty at the cross-section level is evaluated by comparing the theory predictions at the nominal and shifted beam energies. For the inclusive jet measurement, it amounts for 0.2\% at low \pt{} and 0.9\% at high \pt{} in the central region and rises to 2\% at the highest \pt and high rapidity. In the dijet measurement, this uncertainty is 0.2\% at low $\twomass{j}{j}$ and 0.8\% at high $\twomass{j}{j}$ in the first \ystar{} bin and reaches 1\% at the highest $\twomass{j}{j}$ and in the last \ystar{} bin.

In order to assess the statistical precision of the systematic uncertainty estimates, each component is re-evaluated using a set of pseudo-experiments. The statistical fluctuations of the systematic uncertainty estimates are minimised using a smoothing procedure. To achieve this, for each component, the \pt{} ($\twomass{j}{j}$) bins are combined until the propagated uncertainty value in the bin has a Gaussian statistical significance larger than two standard deviations. A Gaussian kernel smoothing~\cite{PERF-2012-01} is used to obtain the values in the original fine bins.

\begin{figure}[htp!]
	\begin{center}
                \subfigure[inclusive jet]{\includegraphics[width=0.45\linewidth]{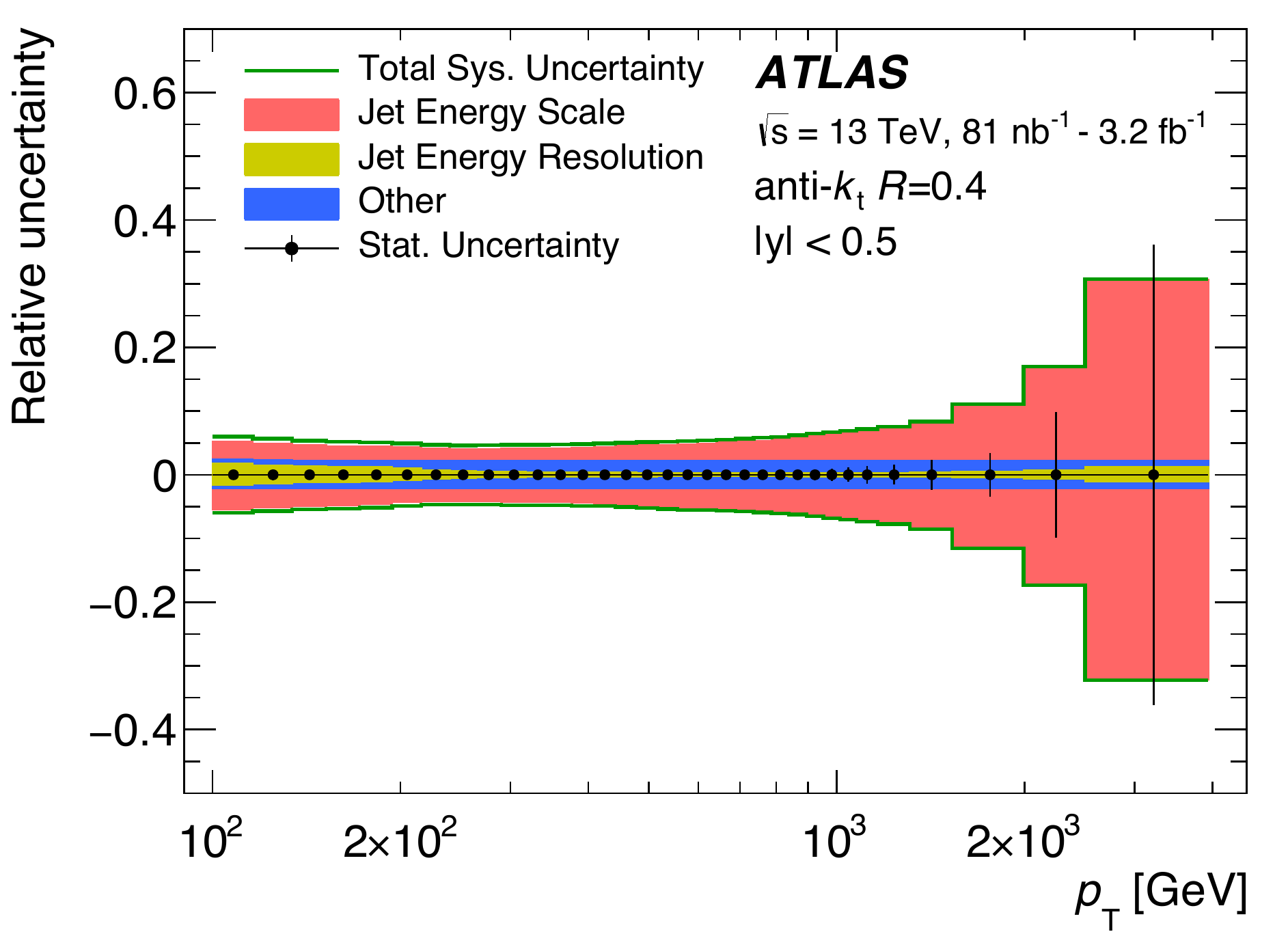}}
                \subfigure[inclusive jet]{\includegraphics[width=0.45\linewidth]{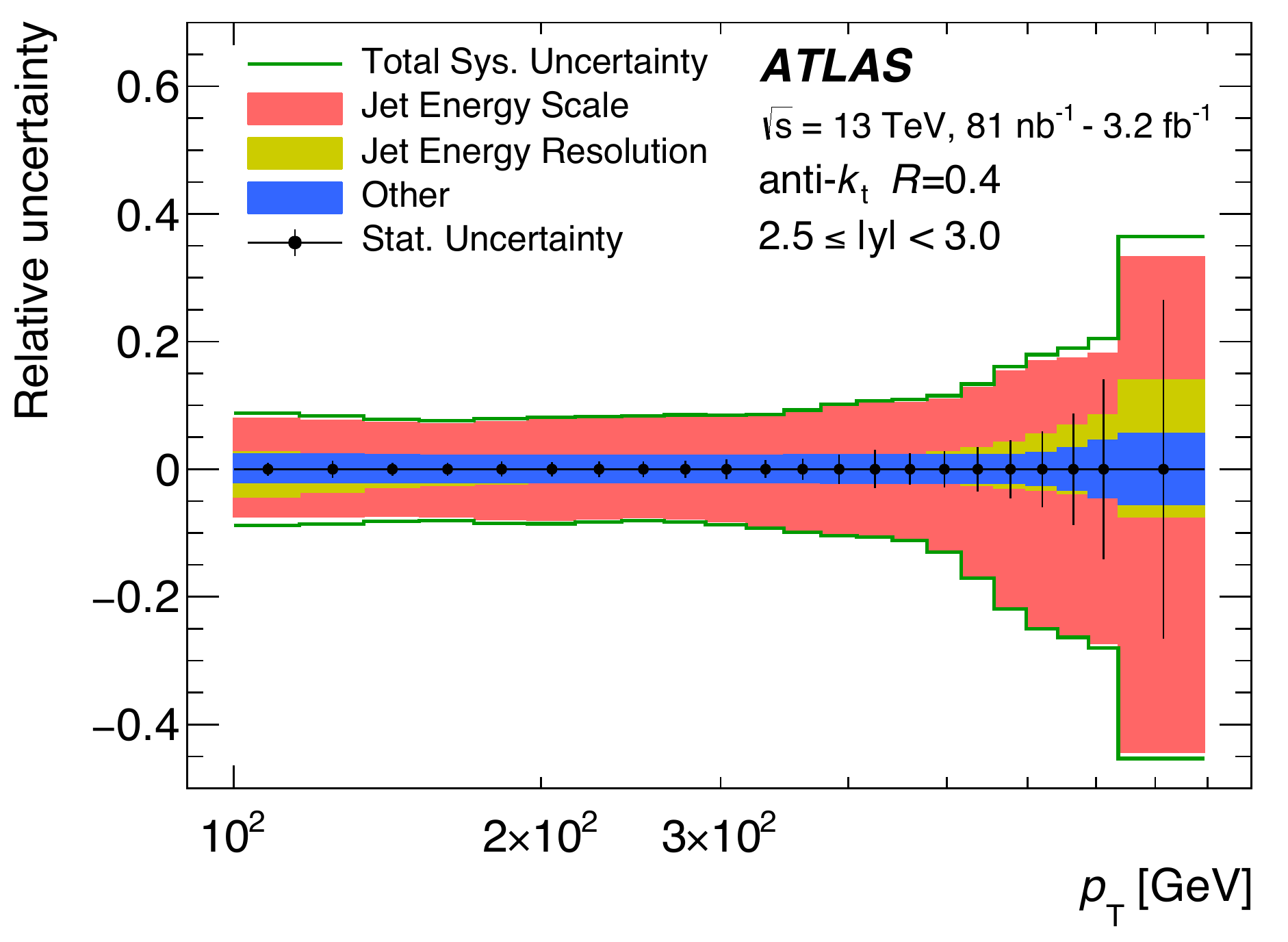}}\\
                \subfigure[dijet]{\includegraphics[width=0.45\linewidth]{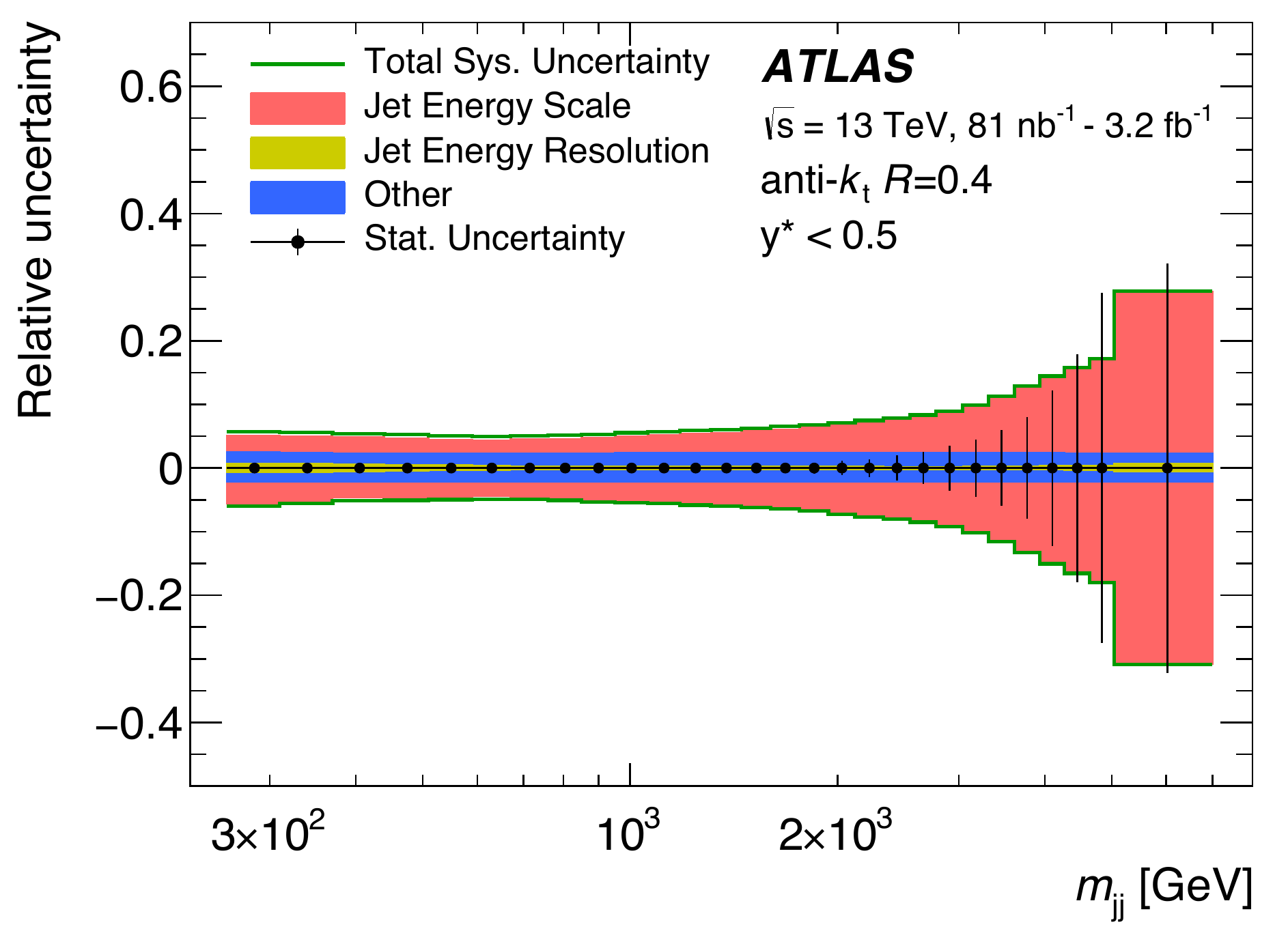}}
                \subfigure[dijet]{\includegraphics[width=0.45\linewidth]{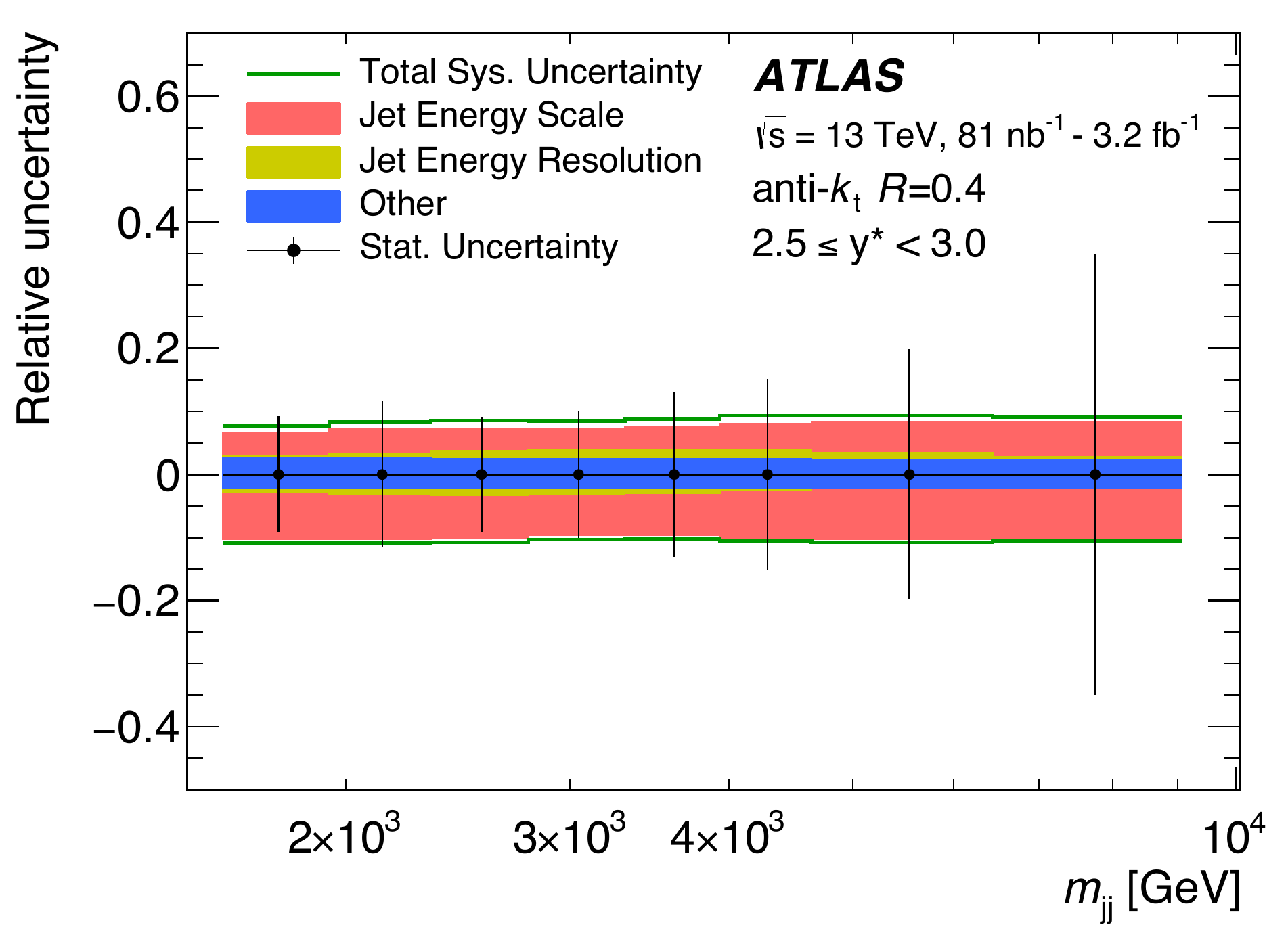}}
		\caption{Relative systematic uncertainty for the inclusive jet cross-section as a function of the jet \pt{} for the first and last rapidity bins ((a) and (b) respectively) and for the dijet cross-section as a function of $\twomass{j}{j}$ for the first and last \ystar{} bins ((c) and (d) respectively). The individual uncertainties are shown in different colours: the jet energy scale, jet energy resolution and the other uncertainties (jet cleaning, luminosity and unfolding bias). The total systematic uncertainty, calculated by adding the individual uncertainties in quadrature, is shown as a green line. The statistical uncertainty is shown as vertical black lines.\label{fig:SystUncert}}
	\end{center}
\end{figure}

Figure~\ref{fig:SystUncert} shows the individual components of the systematic uncertainties added in quadrature for the inclusive jet and dijet cross-section measurements in representative phase-space regions. In the central (forward) region the total uncertainty in the inclusive jet measurement is about 5\% (8\%) at medium \pt{} of 300\texttt{--}600\,\GeV. The uncertainty increases towards both lower and higher pT reaching 6\% (10\%) at low \pt{} and 30\% ([-45\%,+40\%]) at high \pt{}.

The total uncertainty in the dijet measurement is about 5\% (10\%) at medium $\twomass{j}{j}$ of 500\texttt{--}1000\,\GeV{} (2000\texttt{--}3000\,\GeV) in the first (last) \ystar{} bin. The uncertainty increases towards both lower and higher $\twomass{j}{j}$ reaching 6\% at low $\twomass{j}{j}$ and 30\% at high $\twomass{j}{j}$ in the first \ystar{} bin. In the last \ystar{} bin no significant dependence on $\twomass{j}{j}$ is observed.

\FloatBarrier

\section{Theoretical predictions}
\label{sec:theory}
Theoretical predictions of the cross-sections are obtained using NLO and NNLO pQCD calculations with corrections for non-perturbative and electroweak effects.

\subsection{Next-to-leading-order pQCD calculations}

The NLO pQCD predictions are calculated using NLOJET++ 4.1.3~\cite{Nagy:2003tz} interfaced to APPLGRID~\cite{applgrid:2009}
for fast and flexible calculations with various PDF sets and various values of the renormalisation and
factorisation scales. The inclusive jet cross-section prediction is calculated using $\pt^\text{max}$, the transverse momentum of the leading jet in the event, as the renormalisation scale, $\mu_\text{R}$, and the factorisation scale, $\mu_\text{F}$. An alternative scale choice, $\mu_\text{R}=\mu_\text{F}=p_{\rm T}^{\mathrm{jet}}$, the \pt{} of each individual jet that enters the cross-section calculation, is also considered. This scale choice is proposed in Ref.~\cite{ExtraScale}. Both scale choices were used in the previous ATLAS analysis at $\sqrt{s}=8$\,\TeV~\cite{STDM-2015-01}. 
For the dijet cross-section calculation the scale choice is $\mu_\text{R}=\mu_\text{F}=\pt^\text{max}\exp(0.3\ystar)$, as suggested in Ref.~\cite{Ellis92} and previously used in the ATLAS dijet analysis at 7\,\TeV~\cite{STDM-2010-03}. The predictions are calculated using several PDFs provided by the LHAPDF6~\cite{LHAPDF6} library: the NLO CT14~\cite{Dulat:2015mca}, MMHT 2014~\cite{Harland-Lang:2014zoa}, NNPDF 3.0~\cite{Ball:2014uwa}, and HERAPDF 2.0~\cite{Abramowicz:2015mha} sets, and the NNLO ABMP16~\cite{ABMP16} set.
The value of the strong coupling constant, $\alpha_\text{s}$, is taken from the corresponding PDF set.

The main uncertainties in the NLO predictions come from uncertainties associated with the PDFs, the choice of renormalisation and factorisation scales, and the uncertainty in the value of $\alpha_\text{s}$.
PDF uncertainties are defined at the 68\% CL and propagated through the calculations following the prescription given for each PDF set, as recommended by the PDF4LHC group for PDF-sensitive analyses~\cite{PDF4LHC2014}. 
Calculations are redone with varied renormalisation and factorisation scales to estimate the uncertainty
due to missing higher-order terms in the pQCD expansion. The nominal scales are independently varied up or down by a factor of two in both directions excluding opposite variations of $\mu_\text{R}$ and $\mu_\text{F}$. The envelope of resulting variations of the
prediction is taken as the scale uncertainty. The difference between the predictions obtained with the $\pt^{\mathrm{max}}$ and the $p_{\rm T}^{\mathrm{jet}}$ scale choice is treated as an additional uncertainty. The uncertainty from $\alpha_\text{s}$ is evaluated by calculating the
cross-sections using two PDF sets that differ only in the value of $\alpha_\text{s}$ used and then scaling the cross-section difference corresponding to an $\alpha_\text{s}$ uncertainty $\Delta \alpha_\text{s}=0.0015$ as recommended in Ref.~\cite{PDF4LHC2014}.

\begin{figure}[htp!]
        \begin{center}
                \subfigure[inclusive jet]{\includegraphics[width=0.45\linewidth]{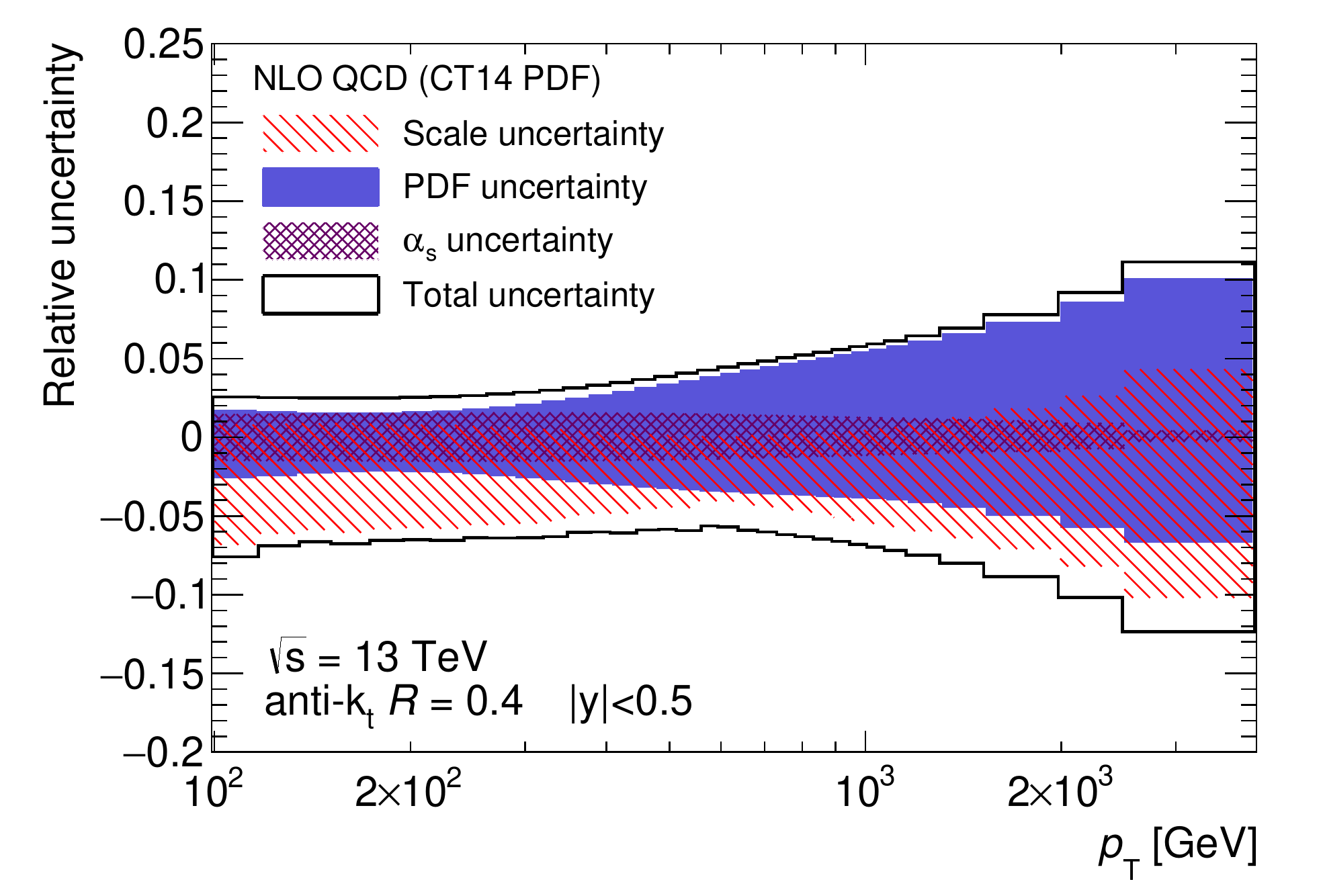}}
                \subfigure[inclusive jet]{\includegraphics[width=0.45\linewidth]{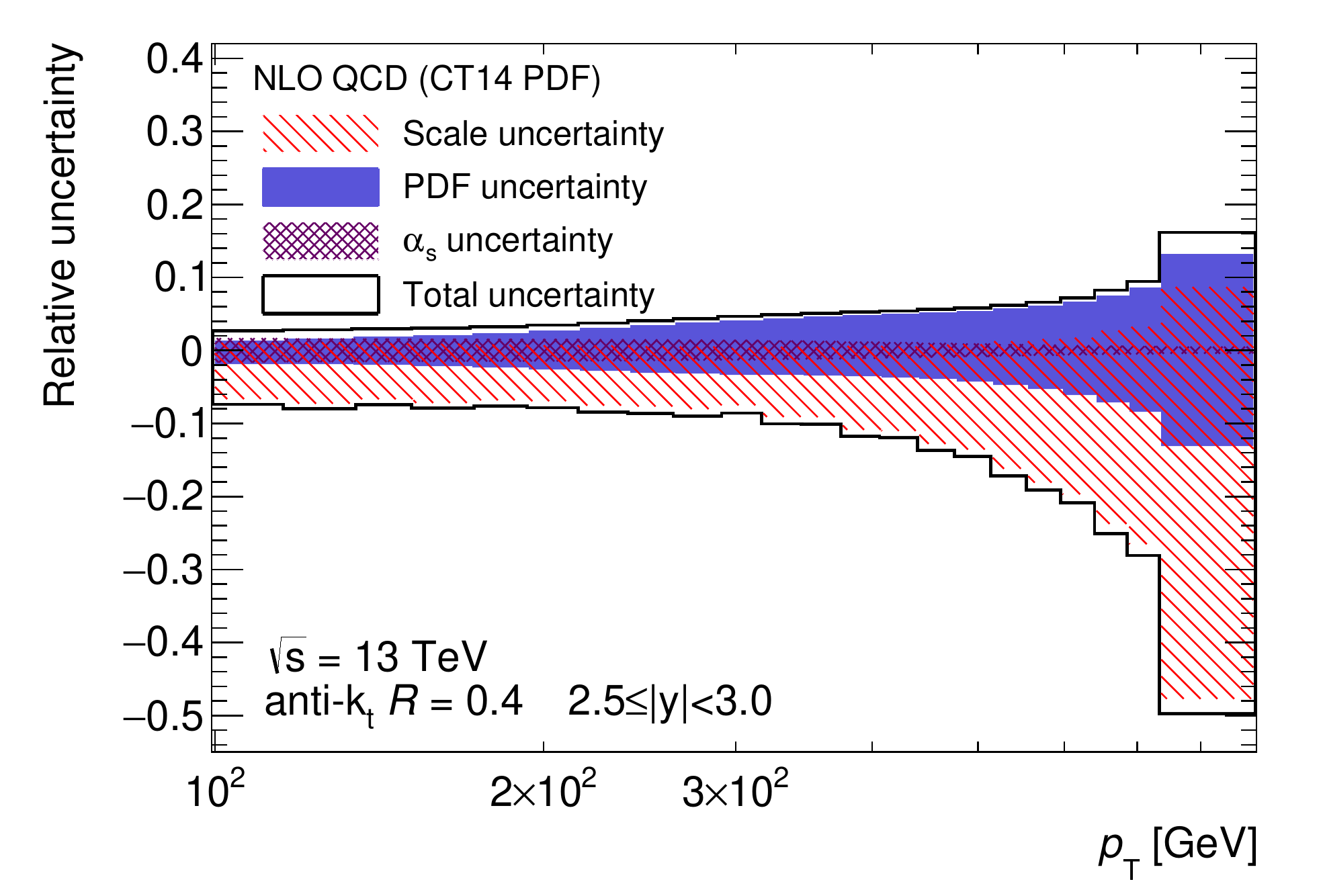}}\\
                \subfigure[dijet]{\includegraphics[width=0.45\linewidth]{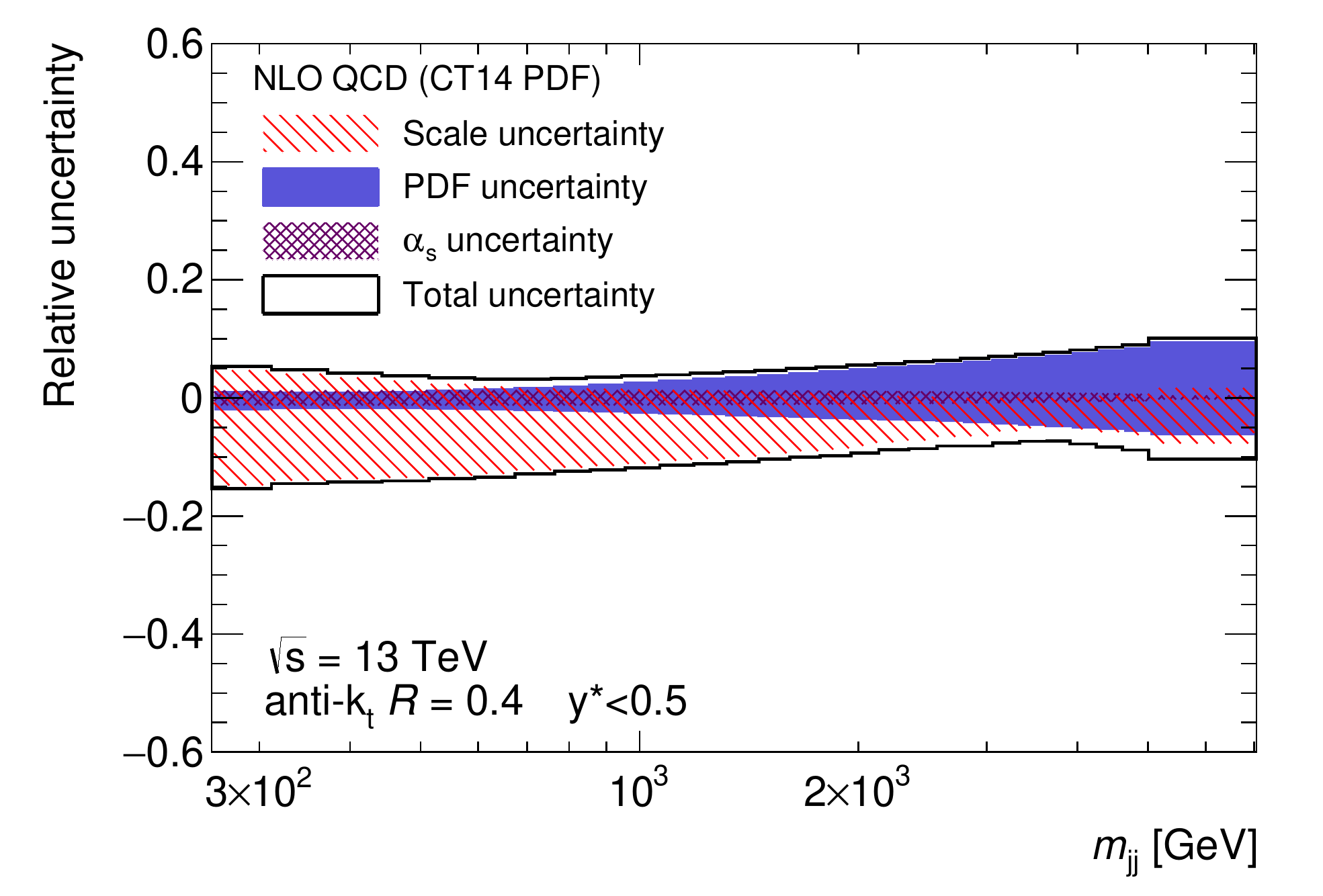}}
                \subfigure[dijet]{\includegraphics[width=0.45\linewidth]{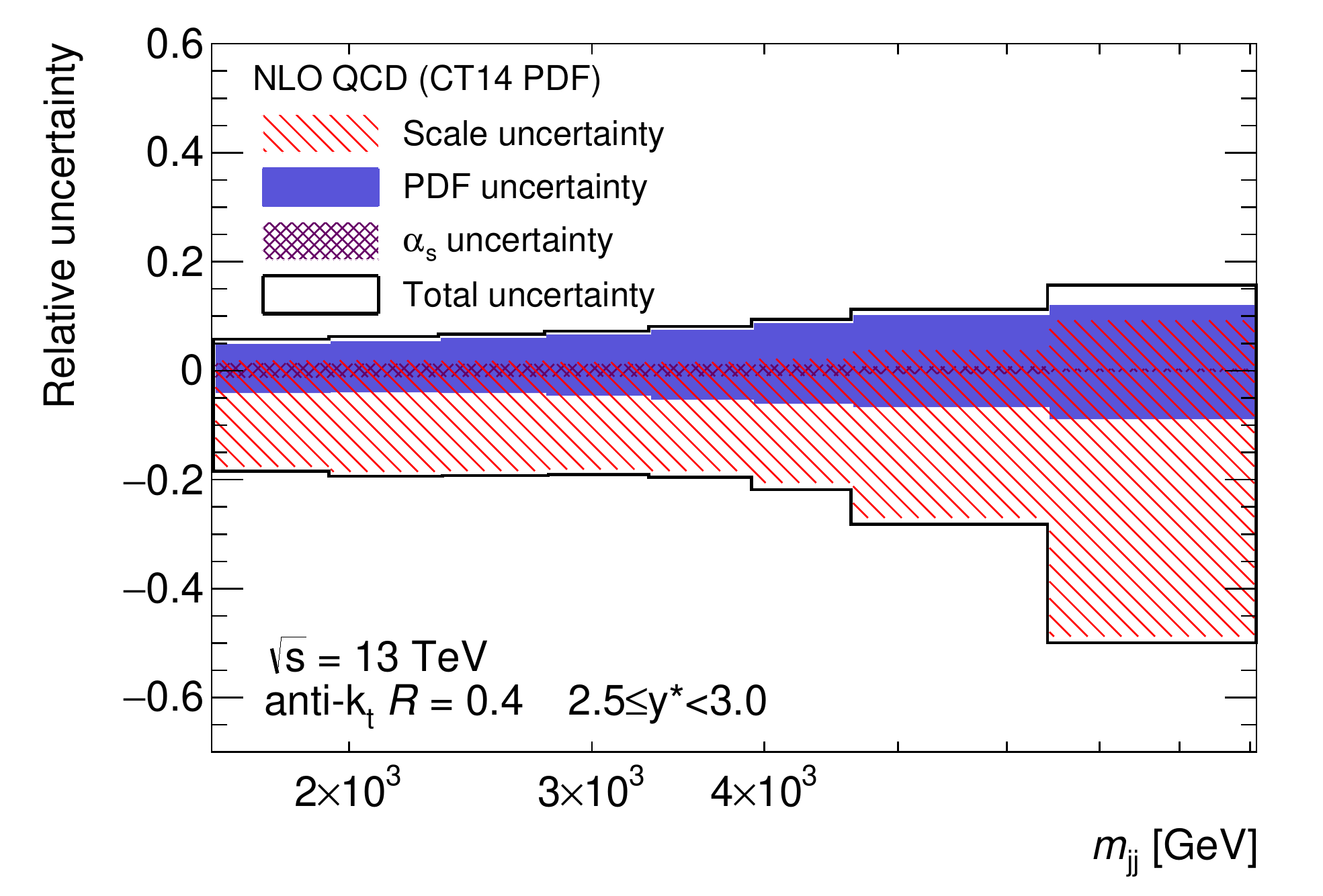}}
		\caption{Relative NLO QCD uncertainties in the jet cross-sections calculated using the CT14 PDF set. Panels a,b (c,d) correspond respectively to the first and last $|y|$ (\ystar) bins for the inclusive jet (dijet) measurement. The uncertainties due to the renormalisation and factorisation scale, the $\alpha_\text{s}$, the PDF and the total uncertainty are shown. The total uncertainty, calculated by adding the individual uncertainties in quadrature, is shown as a black line. 
		\label{fig:NLOuncert}}
        \end{center}
\end{figure}

The uncertainties in the NLO QCD cross-section predictions obtained with the CT14 PDF set are shown in Figure~\ref{fig:NLOuncert} for representative phase-space regions. The uncertainty due to the choice of renormalisation and factorisation scale is dominant in most phase-space regions, rising from 10\% (20\%) at about $\pt = 100$\, \GeV{} ($\twomass{j}{j} = 300$\, \GeV) in the central rapidity (\ystar) bin to about 50\% in the highest \pt ($\twomass{j}{j}$) bins in the most forward rapidity (large \ystar) region.
The PDF uncertainties vary from 2\% to 12\% depending on the jet \pt and rapidity ($\twomass{j}{j}$ and \ystar). The contribution from the $\alpha_\text{s}$ uncertainty is about 2\% at low \pt ($\twomass{j}{j}$) and negligible for the highest \pt ($\twomass{j}{j}$) bin in each rapidity (\ystar) range.

\subsection{Non-perturbative corrections}\label{sec:npc}
Non-perturbative corrections are applied to the parton-level cross-sections from the NLO pQCD calculations. The correction factors are calculated using LO MC event generators, as the bin-by-bin ratio of the nominal particle-level MC cross-sections to the MC cross-section derived from the partons remaining after showering, when the modelling of hadronisation and the underlying event are switched off.
The correction factors are evaluated using several event generators and tunes, which are listed in Table~\ref{tab:MCtunes}. The baseline correction is taken from \pythiaeight{} using the A14 tune with the NNPDF2.3 LO PDF set. The envelope of all corrections is considered as a systematic uncertainty.

\begin{table}[htp!]
\scriptsize
\centering
\tabcolsep=0.08cm
\renewcommand{\arraystretch}{1.2}
\begin{tabular}{l|ccccccc}
\hline\hline
& CTEQ6L1~\cite{Pumplin:2002vw} & CTEQ6L1~\cite{Pumplin:2002vw} & MSTW2008LO~\cite{Martin:2009iq} & CT10 & NNPDF2.3LO & NNPDF2.3LO & CTEQ6L1~\cite{Pumplin:2002vw} \\
\hline
\pythiaeight & 4C~\cite{Corke:2010yf} & AU2~\cite{ATLAS-PHYS-CONF-2012-003} & A14~\cite{ATL-PHYS-PUB-2014-021} & AU2~\cite{ATLAS-PHYS-CONF-2012-003} & MONASH~\cite{Skands:2014pea} & A14~\cite{ATL-PHYS-PUB-2014-021} & A14~\cite{ATL-PHYS-PUB-2014-021} \\
\hline
\herwigpp    & UE-EE-5~\cite{Gieseke:2012ft,Seymour:2013qka} & UE-EE-4~\cite{Gieseke:2012ft,Seymour:2013qka} & UE-EE-5~\cite{Gieseke:2012ft,Seymour:2013qka} & & & & \\
\hline\hline
\end{tabular}
\caption{Summary of the soft-physics model tunes used for the evaluation of the non-perturbative corrections for each event generator and PDF set.
}
\label{tab:MCtunes}
\end{table}

\begin{figure}[htp!]
        \begin{center}
                \subfigure[inclusive jet]{\includegraphics[width=0.45\linewidth]{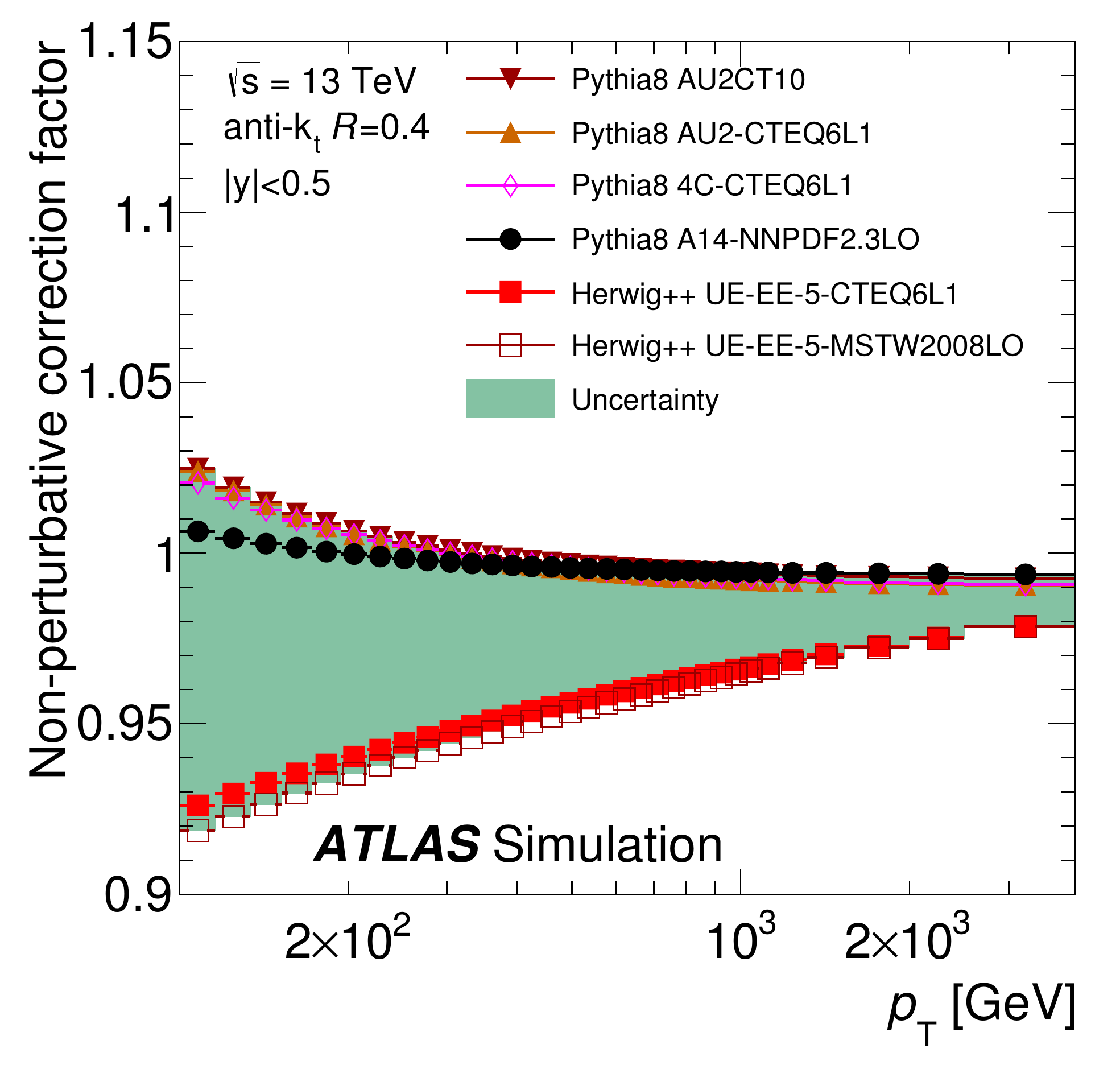}}
                \subfigure[inclusive jet]{\includegraphics[width=0.45\linewidth]{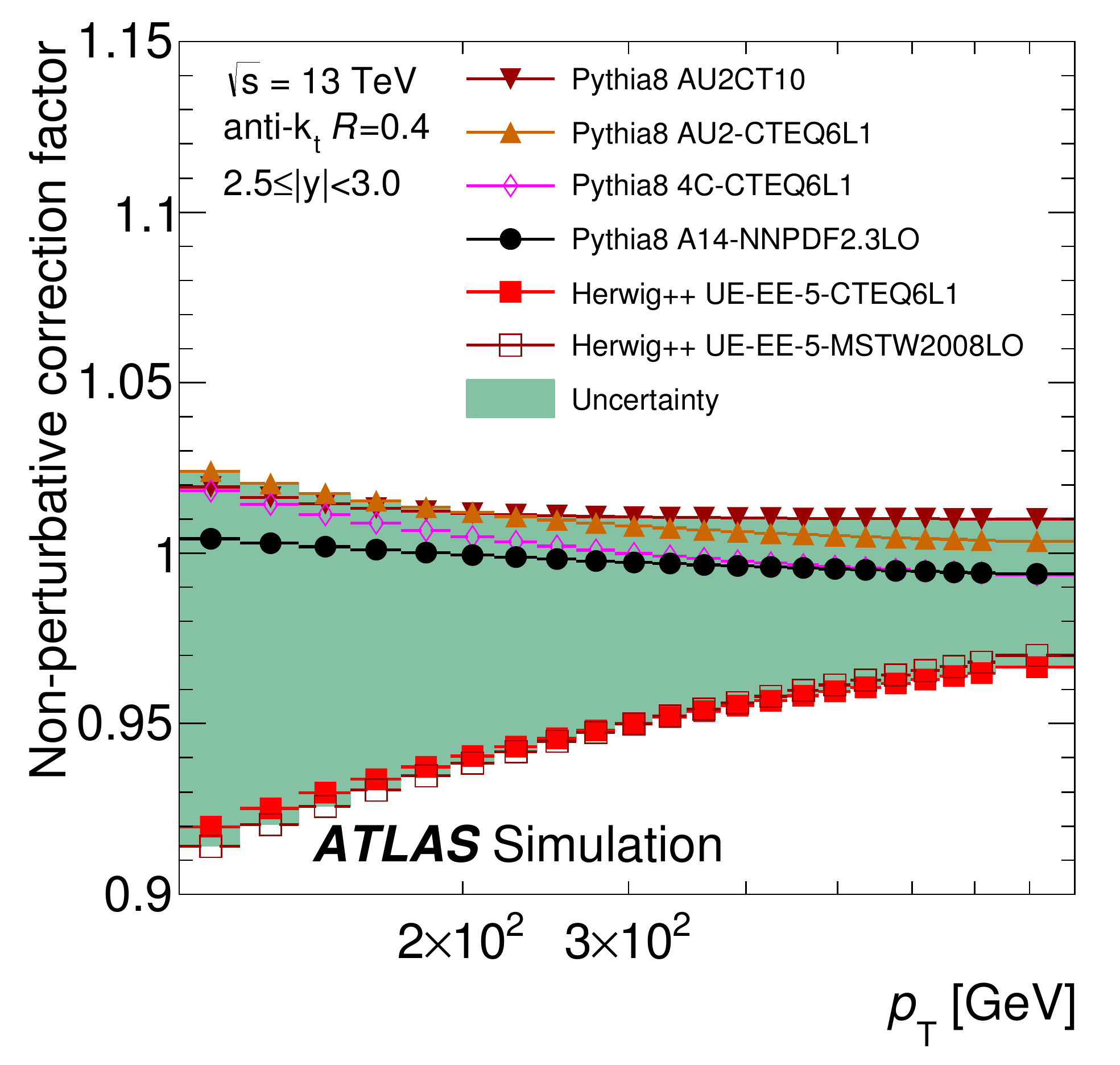}}\\
                \subfigure[dijet]{\includegraphics[width=0.45\linewidth]{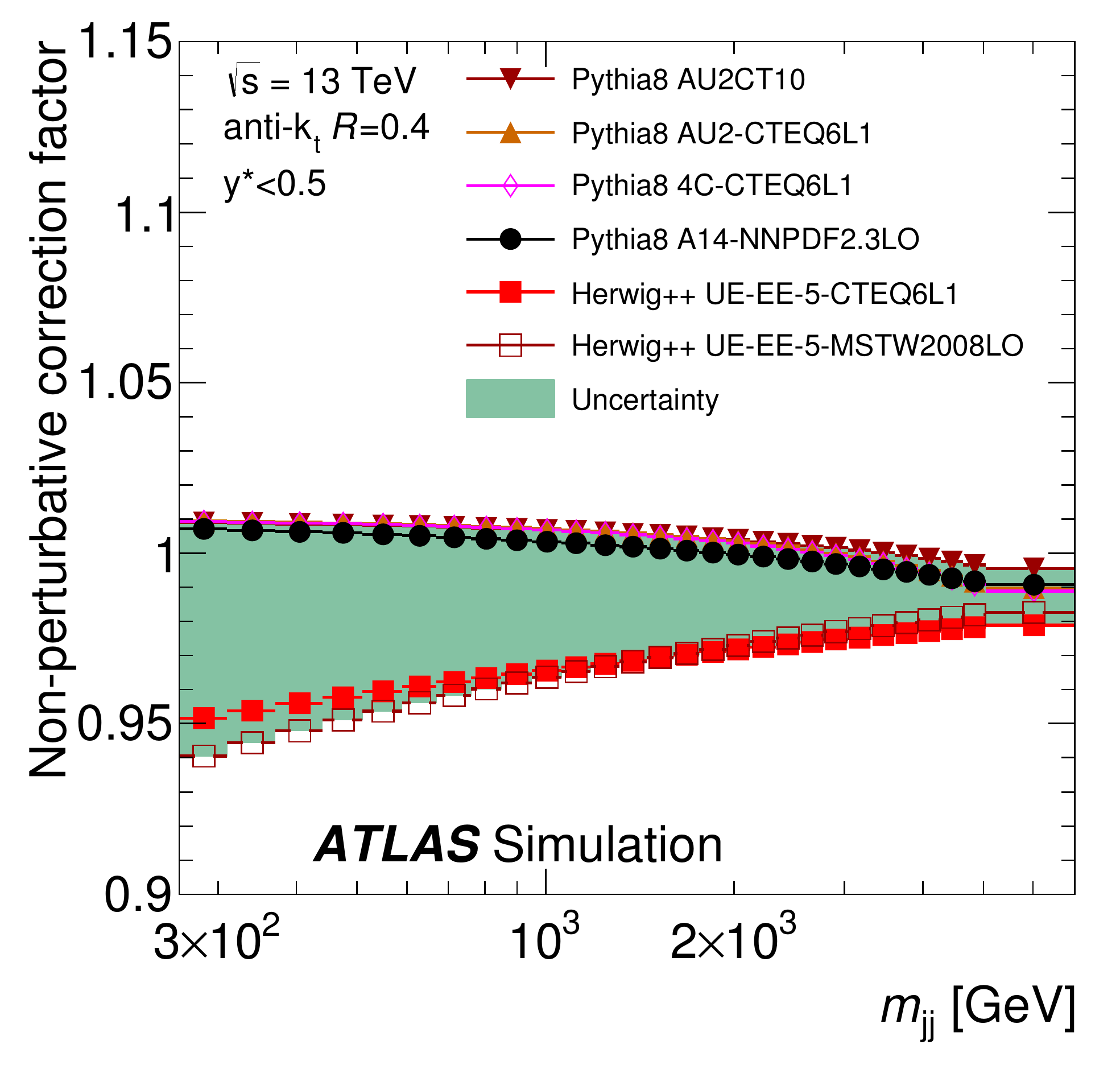}}
                \subfigure[dijet]{\includegraphics[width=0.45\linewidth]{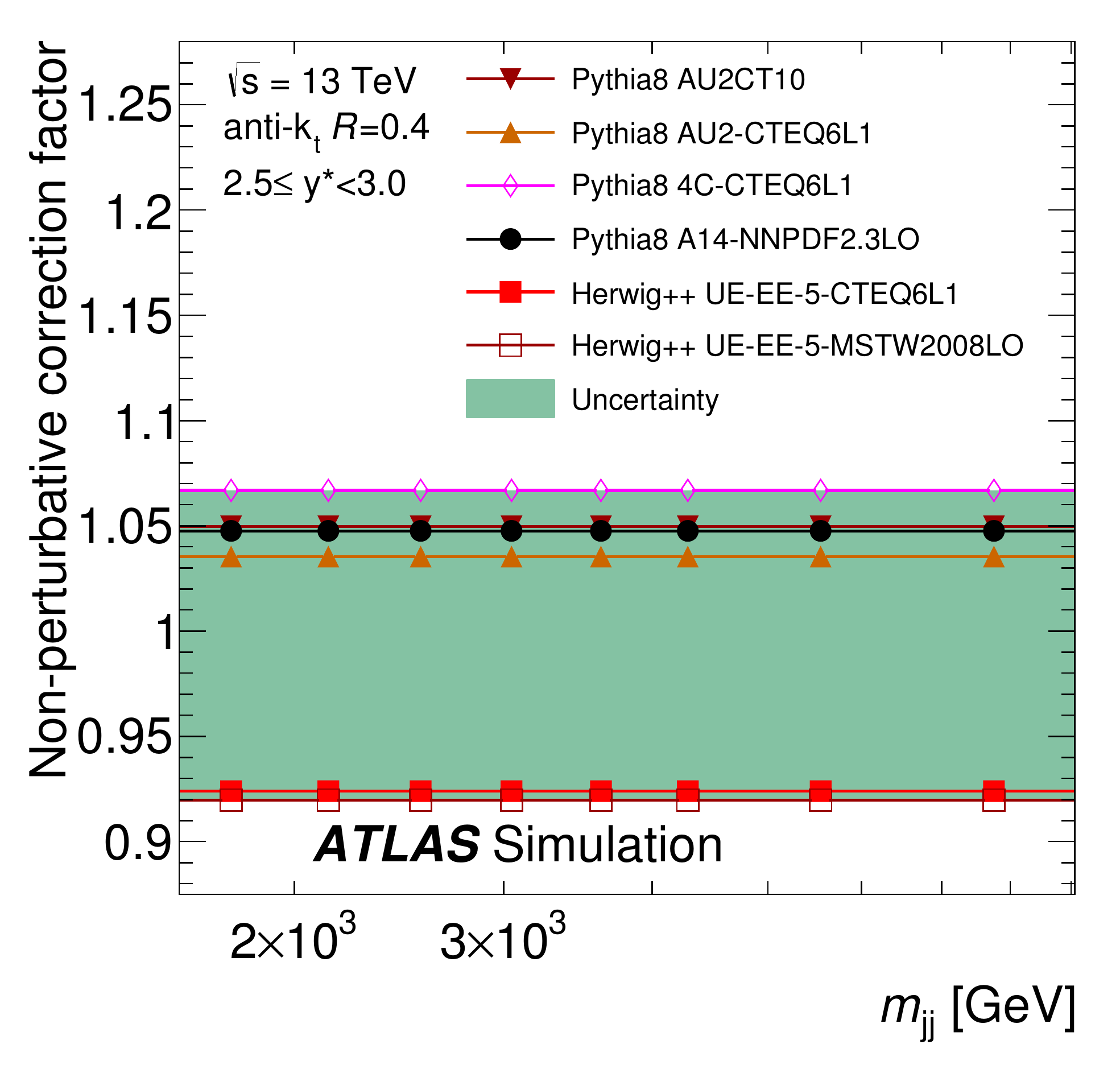}}
		\caption{Non-perturbative correction factors for the (inclusive jet, dijet) NLO pQCD prediction as a function of (jet \pt{}, $\twomass{j}{j}$) for ((a),(c)) the first (rapidity, \ystar) bin and for ((b),(d)) the last (rapidity, \ystar) bin. The corrections are derived using \pythiaeight{} with the A14 tune with the NNPDF2.3 LO PDF set. The envelope of all MC configuration variations is shown as a band.
		\label{fig:NPC}}
        \end{center}
\end{figure}

The correction factors for a set of representative event generators and tunes for the inclusive jet (dijet) cross-section are shown in Figure~\ref{fig:NPC} in illustrative $|y|$ (\ystar) bins as a function of \pt{} ($\twomass{j}{j}$).
The values of the correction are in the range 0.92-1.03 at low \pt{} and 0.98-0.99 (0.97-1.01) at high \pt{} for the first (last) rapidity bin in the inclusive jet measurement, and 0.94-1.01 (0.98-0.99) at low (high) $\twomass{j}{j}$ for the first \ystar{} bin. For the last \ystar{} bin in the dijet measurement, a fixed range 0.92-1.07 is conservatively taken for all $\twomass{j}{j}$ bins due to lack of statistical precision at large $\twomass{j}{j}$. 

\subsection{Electroweak corrections}\label{sec:ew}

The NLO pQCD predictions are corrected for the effects of $\gamma$ and $W^{\pm}/Z$ interactions at tree and one-loop level. They are derived using an NLO calculation of electroweak (EW) contributions to the LO pQCD process. The correction is defined as the ratio of a $2\rightarrow2$ calculation including tree-level effects of order $\alpha_\text{s}^2$, $\alpha^2$, and $\alpha_\text{s}\alpha$ (from interference of QCD and EW diagrams), plus weak loop corrections of order $\alpha_\text{s}^2 \alpha$ to the LO QCD $2\rightarrow2$ calculation.

The correction factors are derived in the phase space considered for the measurements presented here and were provided by the authors of Ref.~\cite{Dittmaier:2012kx}. No uncertainty associated with these corrections is presently estimated. 

\begin{figure}[htp!]
\begin{center}
  {\includegraphics[width=0.45\linewidth]{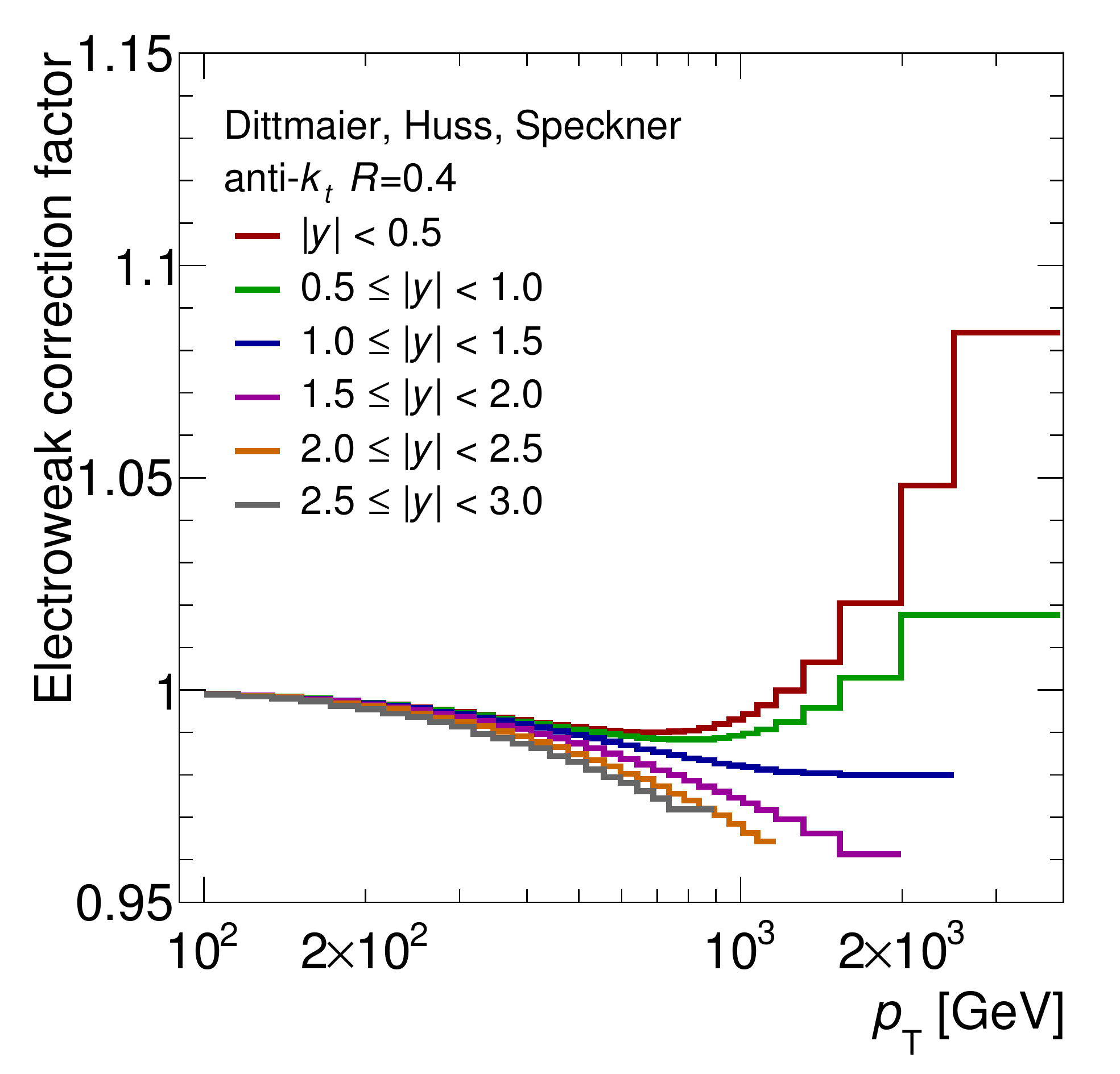}}
  {\includegraphics[width=0.45\linewidth]{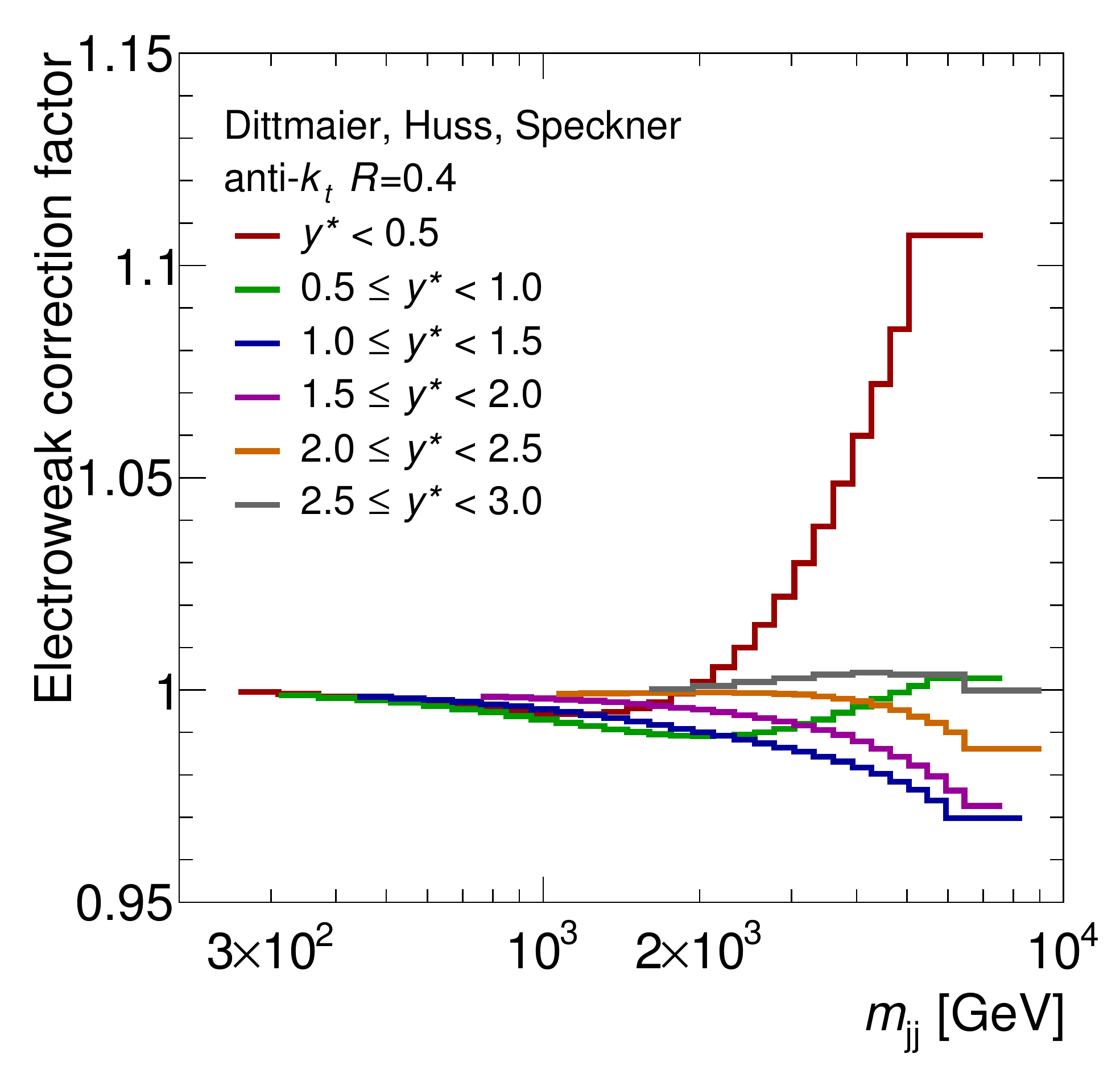}}\\
  \caption{Electroweak correction factors for the inclusive jet (dijet) cross-section as a function of the jet \pT{} ($\twomass{j}{j}$) for all $|y|$ (\ystar) bins.
\label{fig:ew}}
\end{center}
\end{figure}

The electroweak correction factors for the inclusive jet (dijet) cross-section as a function of the jet \pT{} (event $\twomass{j}{j}$) in bins of $|y|$ (\ystar) are shown in Figure~\ref{fig:ew}. The electroweak correction is small for low jet transverse momenta and for low $\twomass{j}{j}$. The correction reaches 8\% at the highest \pT{} (3\,\TeV) for the central $|y|$ bin and is less than 4\% for the rest of the $|y|$ bins. For dijets, the electroweak correction reaches $11$\% at $\twomass{j}{j}=7$\,\TeV{} for the central \ystar{} bin. For the rest of the \ystar{} bins the correction is less than 3\%.

\subsection{Next-to-next-to-leading-order pQCD calculations}

The NNLO pQCD predictions were provided by the authors of Ref.~\cite{NNLO,Currie:2017ctp} using the NNLOJET program and the MMHT 2014 NNLO PDF set for two different choices of the $\mu_\text{R}$ and $\mu_\text{F}$ scales, respectively $p_{\rm T}^{\mathrm{jet}}$ and $p_{\rm T}^{\mathrm{max}}$. The non-perturbative and electroweak corrections described in Sections~\ref{sec:npc} and \ref{sec:ew}, respectively, are applied to the predictions. In addition to the statistical uncertainties on the calculations, which are larger for higher \pt{} and high rapidities, two sources of uncertainty are considered in this NNLO calculation: the scale uncertainty and the systematic uncertainty in the non-perturbative correction. To obtain the scale uncertainty, both scales (renormalisation and factorisation) are varied simultaneously by a factor of 0.5 or 2.~\footnote{A different approach to estimate the scale uncertainty was used for NNLO due to computing time limitations. At NLO the simultaneous variations are not always the dominant ones, although they are at high \pT.}
If both variations yield changes with the same sign, the scale uncertainty is obtained from the larger change.

\FloatBarrier

\section{Results}
\label{sec:results}
\begin{figure}
	\centering
	\includegraphics[width=0.9\textwidth]{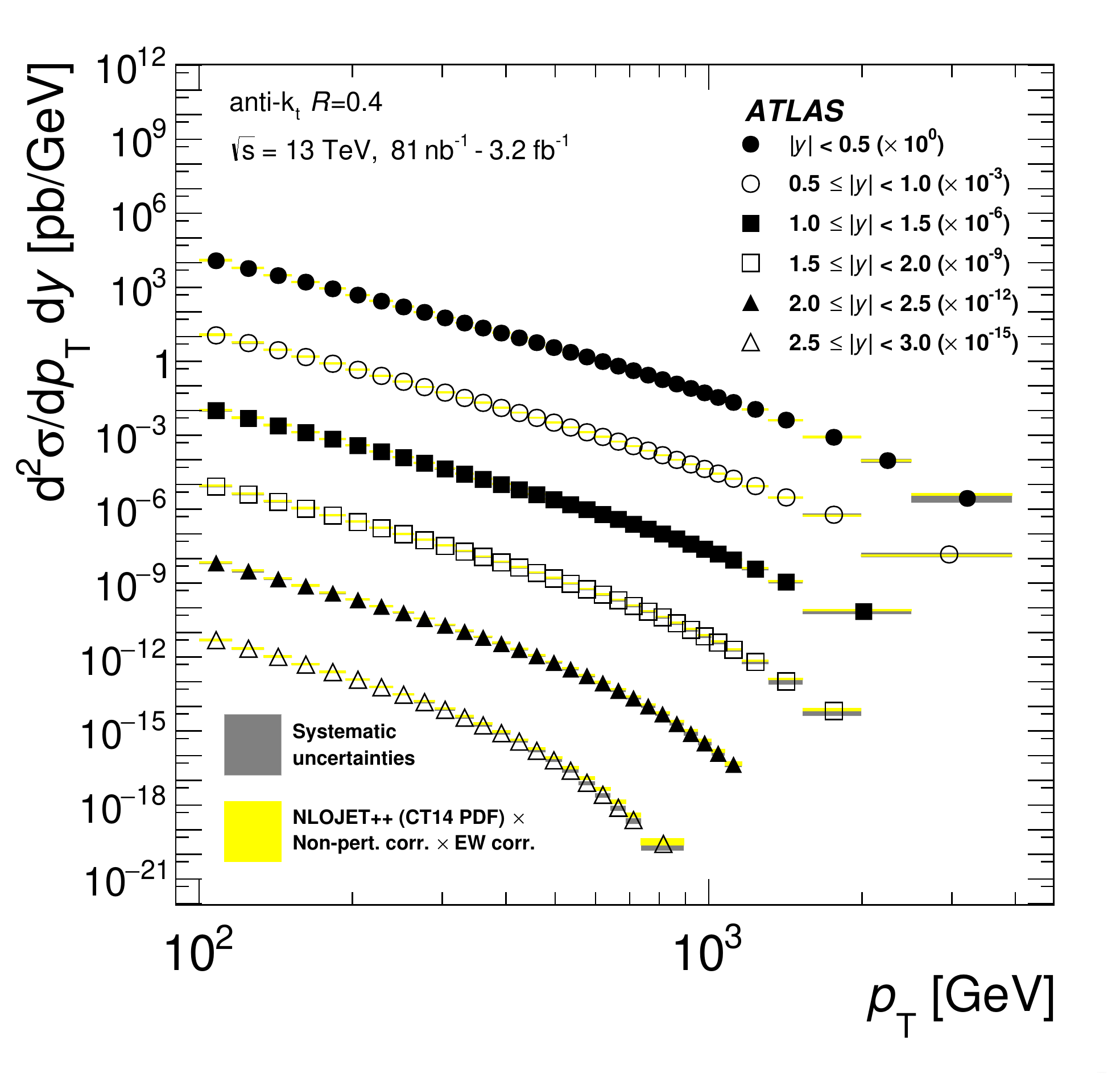}
	\caption{Inclusive jet \xss{} as a function of \pt{} and \absy{},  for \antikt jets with \rfour{}. The statistical uncertainties are smaller than the size of the symbols used to plot the \xs{} values. The dark gray shaded areas indicate the experimental systematic uncertainties. The data are compared to NLO pQCD predictions calculated using NLOJET++ with $p_{\rm T}^{\mathrm{max}}$ as the QCD scale and the CT14~NLO PDF set, to which non-perturbative and electroweak corrections are applied. The light gray (yellow in the online version) shaded areas indicate the predictions with their uncertainties. At low and intermediate \pt{} bins the experimental systematic uncertainties are comparable to the theory uncertainties (drawn on top) and therefore are barely visible.
	\label{fig:InclAbsNLO}}
\end{figure}

\begin{figure}
	\centering
	\includegraphics[width=0.9\textwidth]{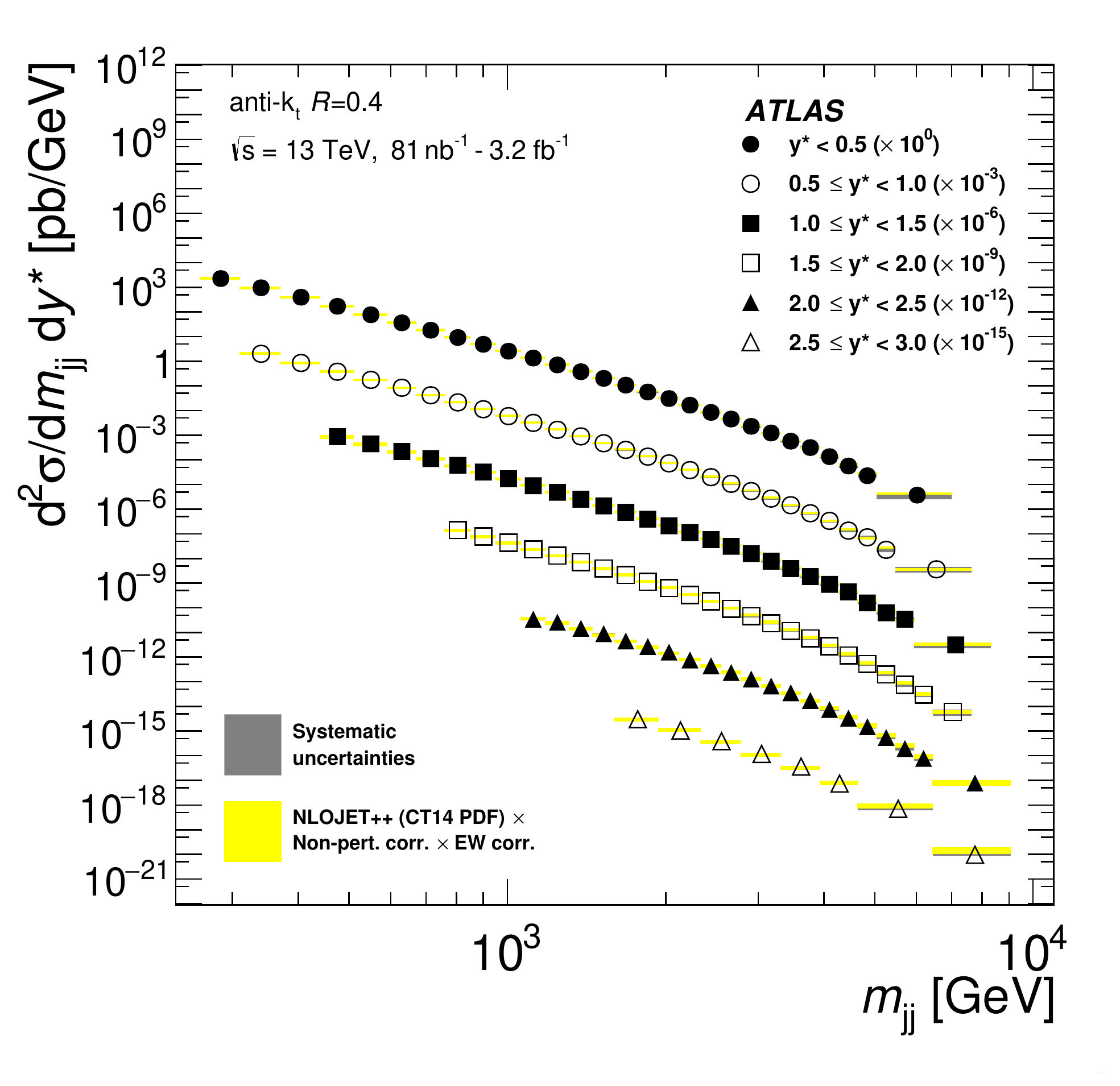}
	\caption{Dijet \xss{} as a function of $\twomass{j}{j}$ and \ystar{},  for \antikt jets with \rfour{}. The statistical uncertainties are smaller than the size of the symbols used to plot the \xs{} values. The dark gray shaded areas indicate the experimental systematic uncertainties. The data are compared to NLO pQCD predictions calculated using NLOJET++ with $\pt^\text{max}\exp(0.3\ystar)$ as the QCD scale and the CT14~NLO PDF set, to which non-perturbative and electroweak corrections are applied. The light gray (yellow in the online version) shaded areas indicate the predictions with their uncertainties. In most $\twomass{j}{j}$ bins the experimental systematic uncertainty is smaller than the theory uncertainties and is therefore not visible. 
	\label{fig:DijetsAbsNLO}}
\end{figure}

The measured double-differential inclusive jet \xss{} are shown in Figure~\ref{fig:InclAbsNLO} as a function of \pt{} for the six jet rapidity bins, and 
the measured double-differential dijet \xss{} are shown in Figure~\ref{fig:DijetsAbsNLO} as a function of $\twomass{j}{j}$ for the six \ystar{} bins. The measurements respectively cover the jet \pt{} range from 100\,\GeV{} to \ptMax{} for $|y|<3.0$, and the $\twomass{j}{j}$ range from \mjjMin{} to \mjjMax{} for $\ystar<3.0$, thus attaining a significantly higher reach than the previous ATLAS measurements~\cite{STDM-2013-11,STDM-2015-01,STDM-2012-03}. The NLO pQCD predictions using the CT14 PDF set corrected for non-perturbative and electroweak effects are also shown in both figures.

\begin{figure}
	\centering
	\includegraphics[width=\linewidth]{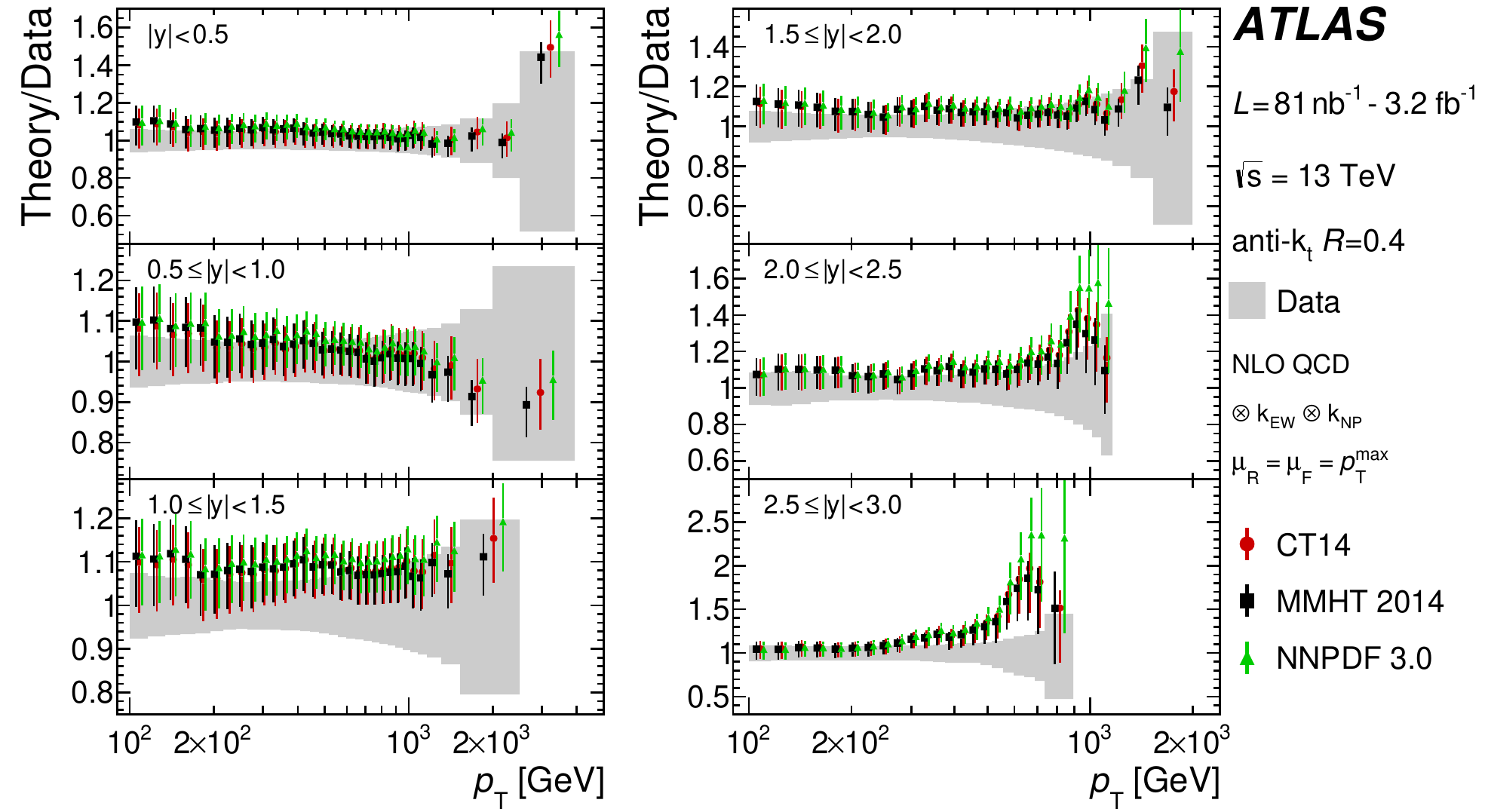}
	\caption{Comparison of the measured inclusive jet \xss{} and the NLO pQCD predictions shown as the ratios of predictions to the measured \xss{}. The ratios are shown as a function of the jet \pt{} in six \absy{} bins for \antikt jets with \rfour{}. The predictions are calculated using NLOJET++ with three different PDF sets (CT14, MMHT 2014, NNPDF 3.0) and non-perturbative and electroweak corrections are applied. The uncertainties of the predictions, shown by the coloured lines, include all the uncertainties discussed in Section~\ref{sec:theory}. The grey bands show the total data uncertainty including both the systematic (JES, JER, unfolding, jet cleaning, luminosity) and statistical uncertainties.} 
	\label{fig:InclXsRatio}
\end{figure}

\begin{figure}
	\centering
	\includegraphics[width=\linewidth]{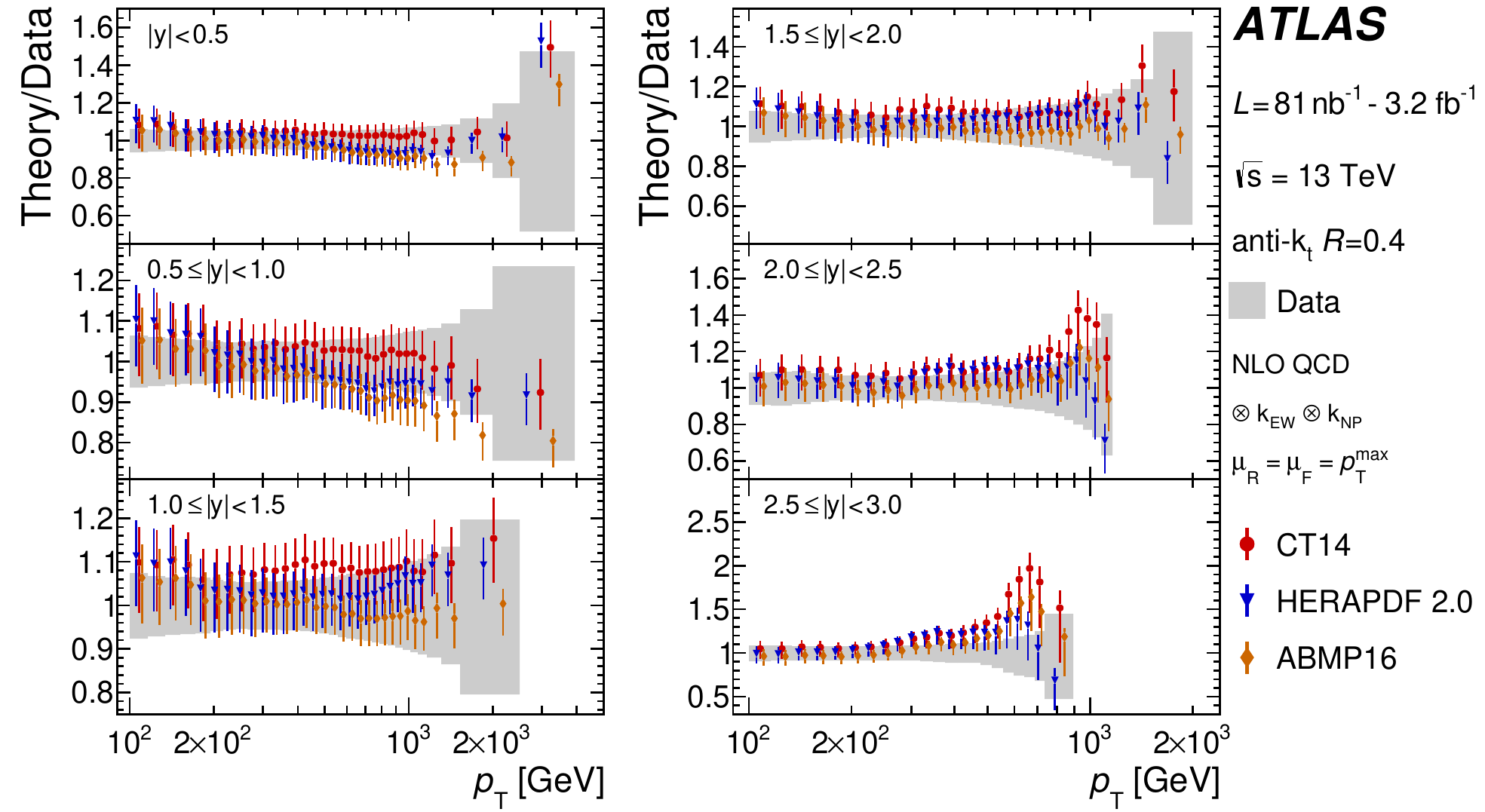}
	\caption{Comparison of the measured inclusive jet \xss{} and the NLO pQCD predictions shown as the ratios of predictions to the measured \xss{}. The ratios are shown as a function of the jet \pt{} in six \absy{} bins for \antikt jets with \rfour{}. The predictions are calculated using NLOJET++ with three different PDF sets (CT14, HERAPDF 2.0, ABMP16) and non-perturbative and electroweak corrections are applied. The uncertainties of the predictions, shown by the coloured lines, include all the uncertainties discussed in Section~\ref{sec:theory}. The grey bands show the total data uncertainty including both the systematic (JES, JER, unfolding, jet cleaning, luminosity) and statistical uncertainties.}
	\label{fig:InclXsRatio_2}
\end{figure}

The ratios of the NLO pQCD predictions to the measured inclusive jet cross-sections as a function of \pt{} in the six jet rapidity bins are shown in Figure~\ref{fig:InclXsRatio} (Figure~\ref{fig:InclXsRatio_2}) for the CT14, MMHT 2014 and NNPDF 3.0 (CT14, ABMP16 and HERAPDF 2.0) PDF sets. The CT14 case is repeated in both figures to serve as a reference for comparison. No significant deviation of the data points from the predictions is seen; the NLO pQCD predictions and data agree within uncertainties. This behaviour is compatible with the results of the comparison between data and the pQCD predictions in the previous ATLAS measurement at $\sqrt{s}=8$\,\TeV~\cite{STDM-2015-01}. In the forward region ($|y|>2$) there is a tendency for the NLO pQCD prediction using the CT14, MMHT 2014 and NNPDF 3.0 PDF sets to overestimate the measured cross-section in the high \pt{} range, although the difference from data does not exceed the range covered by the experimental and theoretical uncertainties.

\begin{figure}
	\centering
	\includegraphics[width=\linewidth]{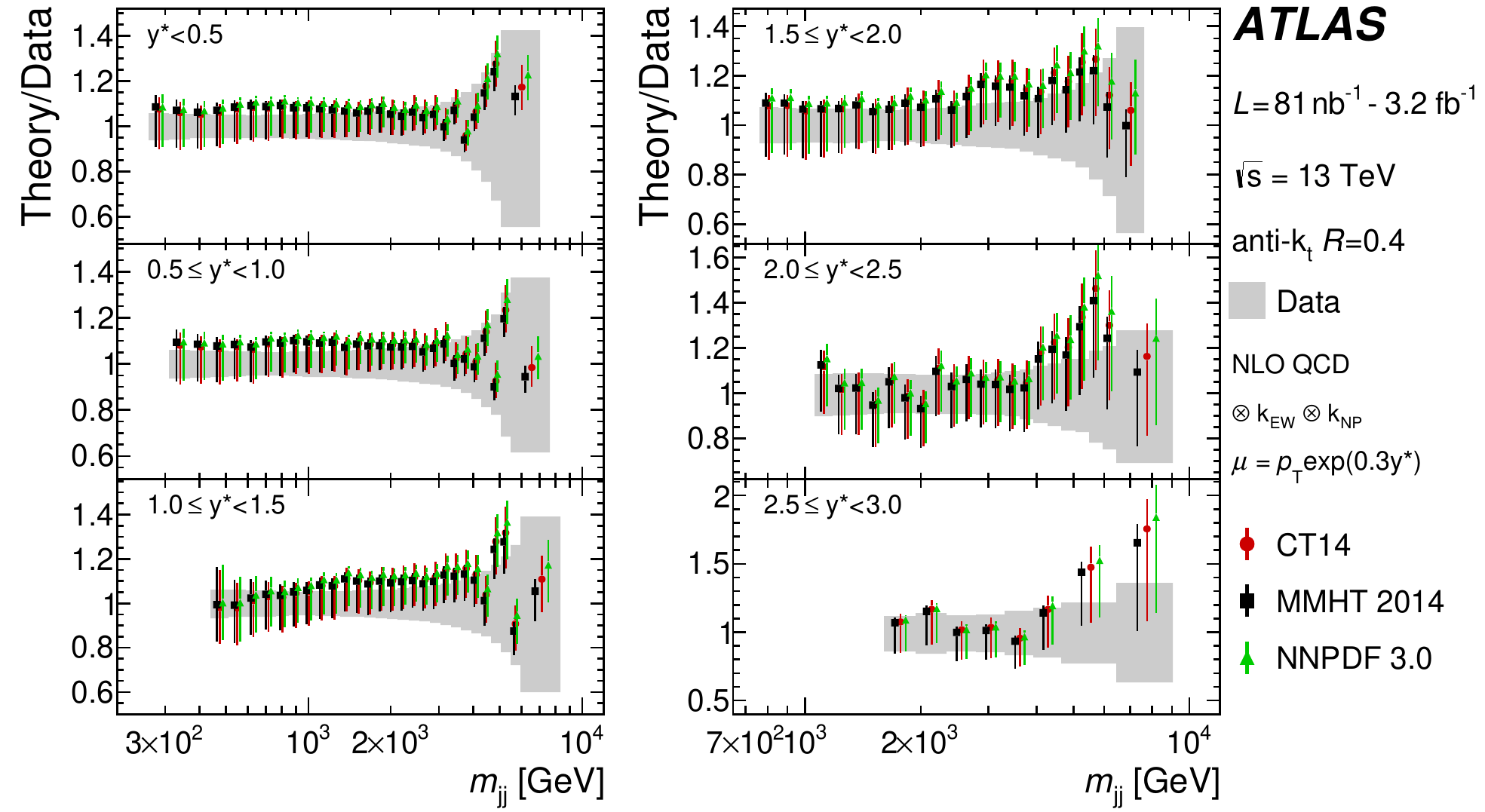}
	\caption{Comparison of the measured dijet \xss{} and the NLO pQCD predictions shown as the ratios of predictions to the measured \xss{}. The ratios are shown as a function of the jet $\twomass{j}{j}$ in six \ystar{} bins for \antikt jets with \rfour{}. The predictions are calculated using NLOJET++ with three different PDF sets (CT14, MMHT 2014, NNPDF 3.0) and non-perturbative and electroweak corrections are applied. The uncertainties of the predictions, shown by the coloured lines, include all the uncertainties discussed in Section~\ref{sec:theory}. The grey bands show the total data uncertainty including both the systematic (JES, JER, unfolding, jet cleaning, luminosity) and statistical uncertainties.}
	\label{fig:DijetXsRatio}
\end{figure}

\begin{figure}
	\centering
	\includegraphics[width=\linewidth]{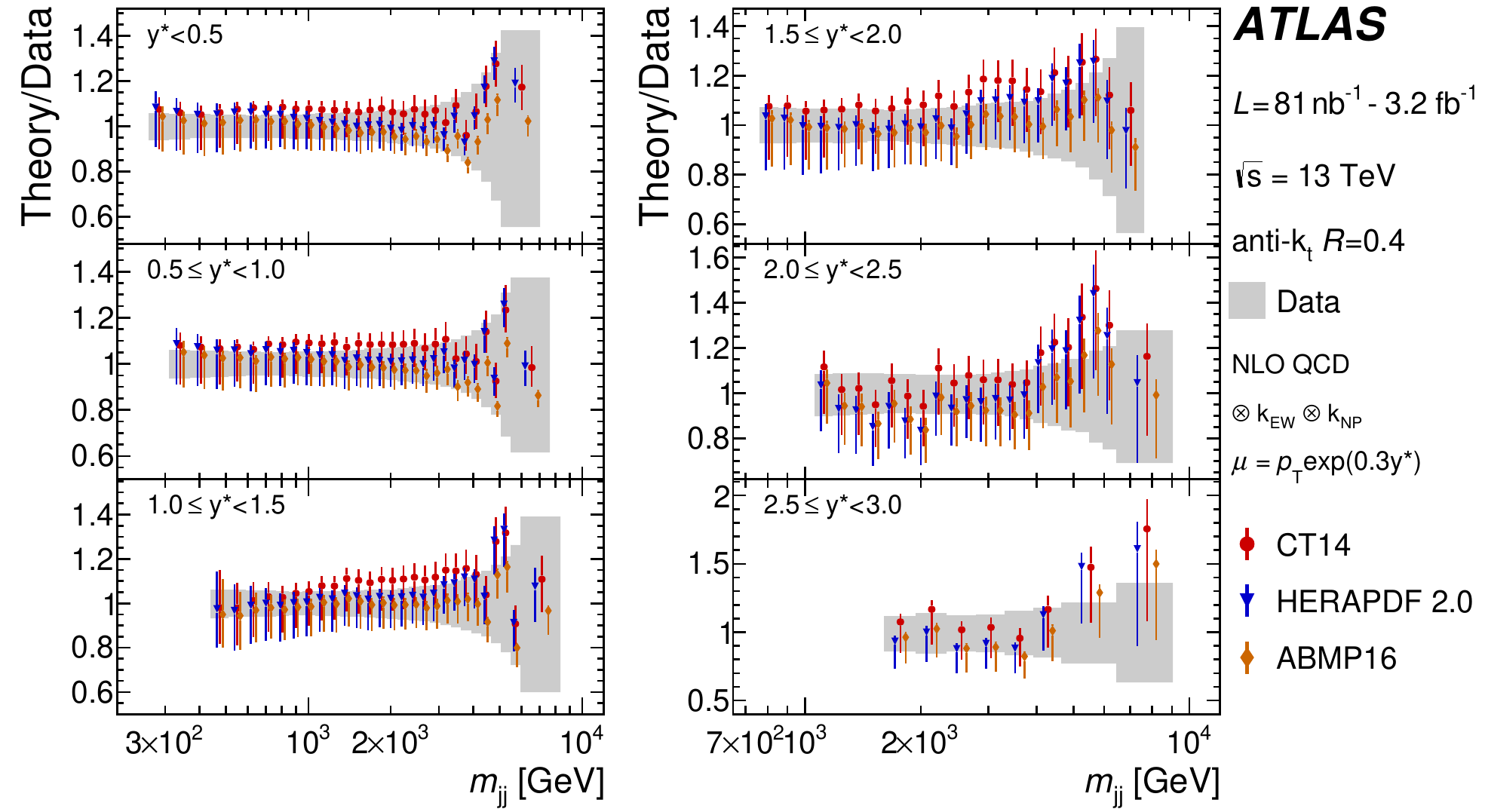}
	\caption{Comparison of the measured dijet \xss{} and the NLO pQCD predictions shown as the ratios of predictions to the measured \xss{}. The ratios are shown as a function of the jet $\twomass{j}{j}$ in six \ystar{} bins for \antikt jets with \rfour{}. The predictions are calculated using NLOJET++ with three different PDF sets (CT14, HERAPDF 2.0, ABMP16) and non-perturbative and electroweak corrections are applied. The uncertainties of the predictions, shown by the coloured lines, include all the uncertainties discussed in Section~\ref{sec:theory}. The grey bands show the total data uncertainty including both the systematic (JES, JER, unfolding, jet cleaning, luminosity) and statistical uncertainties.}
	\label{fig:DijetXsRatio_2}
\end{figure}

The ratios of the NLO pQCD predictions to the measured dijet \xss{} as a function of $\twomass{j}{j}$ in the six \ystar{} bins are shown in Figures~\ref{fig:DijetXsRatio} and \ref{fig:DijetXsRatio_2}. No significant deviation of the data points from the predictions is seen, the NLO pQCD predictions and data agree within uncertainties.

\begin{figure}
	\centering
	\includegraphics[width=\linewidth]{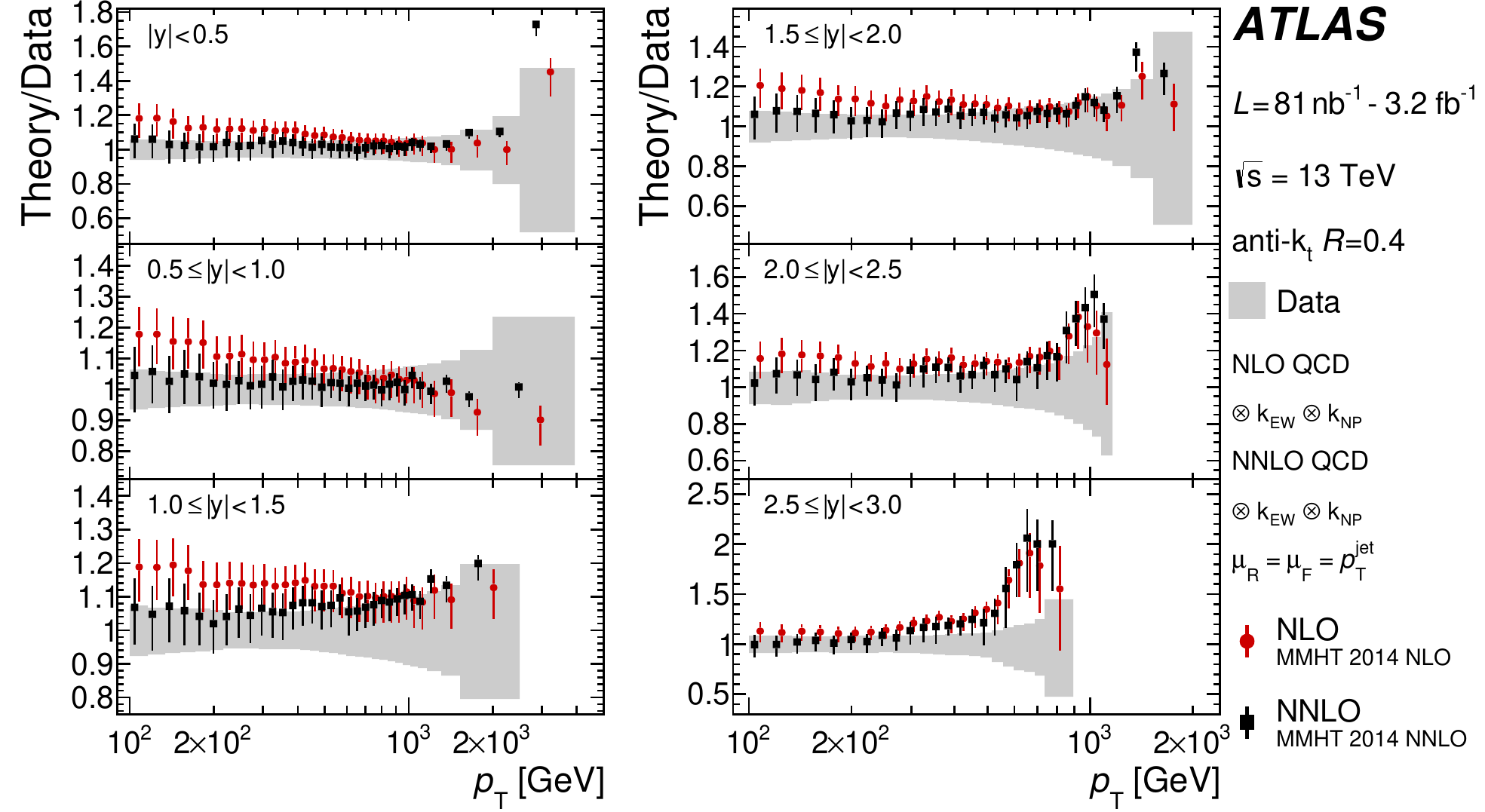}
	\caption{Ratios of the NLO and NNLO pQCD predictions to the measured inclusive jet \xss{}, shown as a function of the jet \pt{} in six \absy{} bins for \antikt jets with \rfour{}. The NLO predictions are calculated using NLOJET++ with the MMHT 2014 NLO PDF set. The NNLO predictions are provided by the authors of Refs.~\cite{NNLO,Currie:2017ctp} using NNLOJET with $p_{\rm T}^{\mathrm{jet}}$ as the QCD scale and the MMHT 2014 NNLO PDF set. Non-perturbative and electroweak corrections are applied to the predictions. The NLO and NNLO uncertainties are shown by the coloured lines, including all the uncertainties discussed in Section~\ref{sec:theory}. The grey bands show the total data uncertainty including both the systematic (JES, JER, unfolding, jet cleaning, luminosity) and statistical uncertainties.}
	\label{fig:InclXsRatio_NNLO_2}
\end{figure}

\begin{figure}
	\centering
	\includegraphics[width=\linewidth]{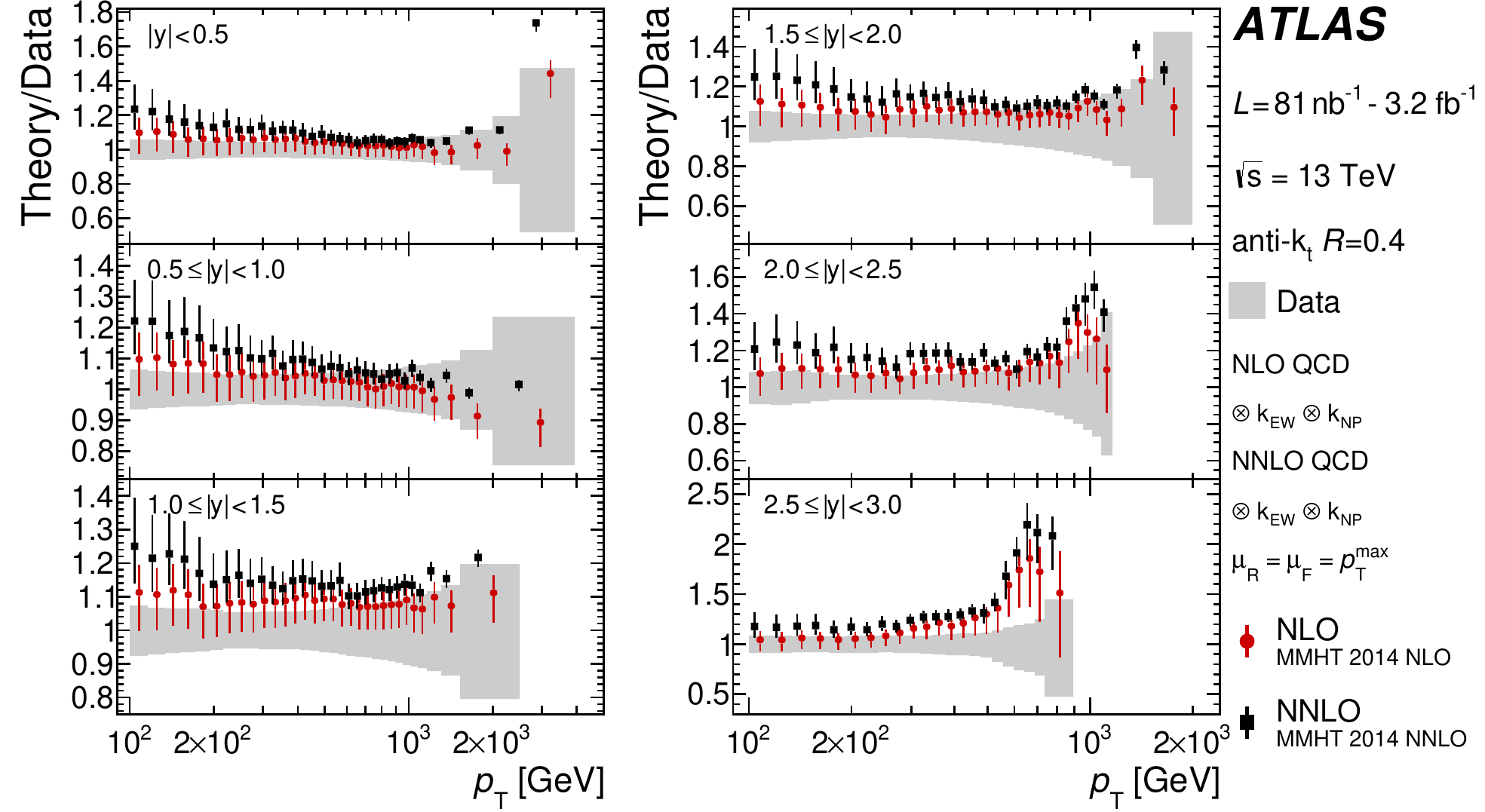}
	\caption{Ratios of the NLO and NNLO pQCD predictions to the measured inclusive jet \xss{}, shown as a function of the jet \pt{} in six \absy{} bins for \antikt jets with \rfour{}. The NLO predictions are calculated using NLOJET++ with the MMHT 2014 NLO PDF set. The NNLO predictions are provided by the authors of Refs.~\cite{NNLO,Currie:2017ctp} using NNLOJET with $p_{\rm T}^{\mathrm{max}}$ as the QCD scale and the MMHT 2014 NNLO PDF set. Non-perturbative and electroweak corrections are applied to the predictions. The NLO and NNLO uncertainties are shown by the coloured lines, including all the uncertainties discussed in Section~\ref{sec:theory}. The grey bands show the total data uncertainty including both the systematic (JES, JER, unfolding, jet cleaning, luminosity) and statistical uncertainties.}
	\label{fig:InclXsRatio_NNLO}
\end{figure}

The ratios of the NNLO pQCD predictions to the measured inclusive jet \xss{} as a function of \pt{} in the six jet rapidity bins are shown in Figures~\ref{fig:InclXsRatio_NNLO_2} and \ref{fig:InclXsRatio_NNLO} for the two different scale choices, respectively $p_{\rm T}^{\mathrm{jet}}$ and $p_{\rm T}^{\mathrm{max}}$, together with the NLO case for comparison. When using $p_{\rm T}^{\mathrm{jet}}$ as a scale, the NNLO pQCD predictions describe the data within uncertainties, with the exception of the forward ($|y|>2$) high \pt{} range where it tends to overestimate the measured cross-section. The predictions using $p_{\rm T}^{\mathrm{max}}$ as the scale overestimate the measured cross-section.

The NLO pQCD predictions, corrected for non-perturbative and electroweak effects, are quantitatively compared to the measurement using the method described in Ref.~\cite{STDM-2012-03}. The $\chi^2$ value and the corresponding observed $p$-value, $P_\text{obs}$, are computed taking into account the asymmetries and the (anti-)correlations of the experimental and theoretical uncertainties. The individual experimental and theoretical uncertainty components are assumed to be uncorrelated among each other and fully correlated across the \pt{} and $|y|$ ($\twomass{j}{j}$ and \ystar{} for dijets) bins. The correlations of the statistical uncertainties across different phase-space regions are taken into account using covariance matrices derived from 1000 pseudo-experiments obtained by fluctuating the data and the MC simulation (see Section~\ref{sec:propunc}).

For the theoretical prediction and separately for each scale choice ($p_{\rm T}^{\mathrm{max}}$ and $p_{\rm T}^{\mathrm{jet}}$), the uncertainties related to the scale variations, the PDF eigenvectors, the non-perturbative corrections and the strong coupling constant are treated as additional uncertainty components. In the case of the NNPDF 3.0 PDF set, the replicas~\cite{Ball:2014uwa} are used to evaluate a covariance matrix, from which the eigenvectors are then determined.

Table~\ref{tab:chi2_IJXS} shows the summary of the observed $P_\text{obs}$ values for each individual rapidity bin of the inclusive jet measurement. 
Table~\ref{tab:chi2_IJXS_global} reports the results obtained from a global fit to all the \pt{} and rapidity bins of the measurement. Given that in this case the observed $P_\text{obs}$ values are very small, the results are presented in terms of the $\chi^2$ per degree of freedom ($\text{dof}$). Table~\ref{tab:chi2_DIJETS} shows the summary of observed $P_\text{obs}$ values for each \ystar{} bin of the dijet measurement, as well as those from a global fit using all the $\twomass{j}{j}$ and \ystar{} bins.

Fair agreement is seen (with $p$-values in the percent range) when considering jet cross-sections in individual jet rapidity or \ystar{} bins treated independently, with some tension present in the 1.5--2.5 rapidity region. Comparable results are obtained for PDF sets determined with similar data. Strong tension between data and theory is observed when considering data points from all jet transverse momentum and rapidity regions in the inclusive jet measurement (Table~\ref{tab:chi2_IJXS_global}), a behaviour already observed in the previous ATLAS measurement at $\sqrt{s}=8$\,\TeV~\cite{STDM-2015-01}. For the dijet measurement, the agreement is fair when considering events from all \ystar{} regions, as observed in the previous ATLAS measurement at $\sqrt{s}=7$\,\TeV~\cite{STDM-2012-03}.

Consideration of all data points together requires a good understanding of the correlations of the experimental and theoretical systematic uncertainties in jet \pt{} and rapidity. Although the correlations of most uncertainties are generally well known, the systematic uncertainties that are based on simple comparisons between two options (two-point uncertainties) are not well defined. This is the case for instance for the in situ multijet balance uncertainties due to different fragmentation models and the theoretical uncertainty related to the alternative scale choice. In these cases, alternative decorrelation scenarios can in principle be used instead of the default full correlation model. In these, systematic uncertainties are split into sub-components whose size varies with jet rapidity and \pt{}, keeping their sum in quadrature equal to the original uncertainty.

Reference~\cite{STDM-2015-01} presents a detailed discussion about the alternative correlation options that can be considered acceptable. The same conclusions are applicable here. Decorrelation scenarios were applied simultaneously to the largest sources of two-point experimental uncertainties (the JES flavour response, the JES multijet $p_\mathrm{T}$-balance fragmentation, and the pile-up energy density in jets) as well as the theoretical uncertainties (the scale variations, the alternative scale choice and the non-perturbative corrections) using the splitting options that yielded the largest $\chi^2$ reduction for each single component in Ref.~\cite{STDM-2015-01}. The $\chi^2$ using the CT14 PDF set and the $p_{\rm T}^{\mathrm{max}}$ scale choice is found to be reduced by 58 units ($\chi^2/\text{dof}=361/177$) compared to the nominal configuration, but the corresponding $p$-value is still $\ll10^{-3}$, in agreement with the conclusions of the previous ATLAS measurement at $\sqrt{s} = 8$\,\TeV~\cite{STDM-2015-01}.

\begin{table}[htp!]
\centering
\tabcolsep=0.2cm
\begin{tabular}{c|rrrrr}
\hline\hline
 & \multicolumn{5}{c}{$P_\text{obs}$} \\
Rapidity ranges & CT14 & MMHT 2014 & NNPDF 3.0 & HERAPDF 2.0 & ABMP16 \\
\hline
$p_{\rm T}^{\mathrm{max}}$ \\
\hline
$|y|<0.5$ & 67\% & 65\% & 62\% & 31\% & 50\% \\
$0.5\le|y|<1.0$ & 5.8\% & 6.3\% & 6.0\% & 3.0\% & 2.0\% \\
$1.0\le|y|<1.5$ & 65\% & 61\% & 67\% & 50\% & 55\% \\
$1.5\le|y|<2.0$ & 0.7\% & 0.8\% & 0.8\% & 0.1\% & 0.4\% \\
$2.0\le|y|<2.5$ & 2.3\% & 2.3\% & 2.8\% & 0.7\% & 1.5\% \\
$2.5\le|y|<3.0$ & 62\% & 71\% & 69\% & 25\% & 55\% \\
\hline
$p_{\rm T}^{\mathrm{jet}}$ \\
\hline
$|y|<0.5$ & 69\% & 67\% & 66\% & 30\% & 46\% \\
$0.5\le|y|<1.0$ & 7.4\% & 8.9\% & 8.6\% & 3.4\% & 2.0\% \\
$1.0\le|y|<1.5$ & 69\% & 62\% & 68\% & 45\% & 54\% \\
$1.5\le|y|<2.0$ & 1.3\% & 1.6\% & 1.4\% & 0.1\% & 0.5\% \\
$2.0\le|y|<2.5$ & 8.7\% & 6.6\% & 7.4\% & 1.0\% & 3.6\% \\
$2.5\le|y|<3.0$ & 65\% & 72\% & 72\% & 28\% & 59\% \\
\hline\hline
\end{tabular}
\caption{Summary of observed $P_\text{obs}$ values from the comparison of the inclusive jet cross-section and the NLO pQCD prediction corrected for non-perturbative and electroweak effects for various PDF sets, for the two scale choices and for each rapidity bin of the measurement.}
\label{tab:chi2_IJXS}
\end{table}

\begin{table}[htp!]
\centering
\tabcolsep=0.15cm
\begin{tabular}{c|c|c|c|c|c}
\hline\hline
$\chi^2/\text{dof}$ & \multirow{2}{*}{CT14} & \multirow{2}{*}{MMHT 2014} & \multirow{2}{*}{NNPDF 3.0} & \multirow{2}{*}{HERAPDF 2.0} & \multirow{2}{*}{ABMP16} \\
all $|y|$ bins &  & & & & \\
\hline
$p_{\rm T}^{\mathrm{max}}$ & 419/177 & 431/177 & 404/177 & 432/177 & 475/177\\
\hline
$p_{\rm T}^{\mathrm{jet}}$ & 399/177 & 405/177 & 384/177 & 428/177 & 455/177\\
\hline\hline
\end{tabular}
\caption{Summary of $\chi^2/\text{dof}$ values obtained from a global fit using all \pt{} and rapidity bins, comparing the inclusive jet cross-section and the NLO pQCD prediction corrected for non-perturbative and electroweak effects for several PDF sets and for the two scale choices. All the corresponding $p$-values are $\ll10^{-3}$.}
\label{tab:chi2_IJXS_global}
\end{table}

\begin{table}[htp!]
\centering
\tabcolsep=0.2cm
\begin{tabular}{c|rrrrr}
\hline\hline
 & \multicolumn{5}{c}{$P_\text{obs}$} \\
\ystar{} ranges & CT14 & MMHT 2014 & NNPDF 3.0 & HERAPDF 2.0 & ABMP16\\
\hline
$y^{*}<0.5$ & 79\% & 59\% & 50\% & 71\% & 71\% \\
$0.5\le y^{*}<1.0$ & 27\% & 23\% & 19\% & 32\% & 31\% \\
$1.0\le y^{*}<1.5$ & 66\% & 55\% & 48\% & 66\% & 69\% \\
$1.5\le y^{*}<2.0$ & 26\% & 26\% & 28\% & 9.9\% & 25\% \\
$2.0\le y^{*}<2.5$ & 41\% & 34\% & 29\% & 3.6\% & 20\% \\
$2.5\le y^{*}<3.0$ & 45\% & 46\% & 40\% & 25\% & 38\% \\
\hline
all \ystar{} bins & 9.4\% & 6.5\% & 11\% & 0.1\% & 5.1\%\\
\hline\hline
\end{tabular}
\caption{Summary of observed $P_\text{obs}$ values obtained from the comparison of the dijet cross-section and the NLO pQCD prediction corrected for non-perturbative and electroweak effects for various PDF sets and for each individual \ystar{} range. The last row of the table corresponds to a global fit using all $\twomass{j}{j}$ and \ystar{} bins of the dijet measurement.}
\label{tab:chi2_DIJETS}
\end{table}

Since the uncertainties in the NNLO pQCD predictions do not yet include the contributions from the PDF and $\alpha_\text{s}$ uncertainties, it is not possible to perform a quantitative comparison to the measurements. However, one can conclude from Figure~\ref{fig:InclXsRatio_NNLO_2} (Figure~\ref{fig:InclXsRatio_NNLO}) that the differences between data and the theoretical predictions at NNLO are smaller (larger) than at NLO for the $p_{\rm T}^{\mathrm{jet}}$ ($p_{\rm T}^{\mathrm{max}}$) scale choice.

\FloatBarrier

\section{Conclusion}
\label{sec:conclusions}
The inclusive jet and dijet \xss{} in proton--proton collisions at $\rts=13$~\TeV{} are measured for jets reconstructed with the \antikt algorithm with a jet radius parameter value of \rfour. The measurements use data collected at the LHC with the ATLAS detector during 2015 corresponding to an integrated luminosity of \LUMI. The inclusive jet \xss{} are measured double-differentially in the jet transverse momentum and jet rapidity in a kinematic region between \ptMin{} and \ptMax{} with $\absy<3$. The dijet \xss{} are measured double-differentially in the invariant mass of the dijet system and half the absolute rapidity separation between the two leading jets with $\absy<3$, covering $\mjjMin<\twomass{j}{j}<\mjjMax$ and $\ystar<3$. The dominant systematic uncertainty arises from the jet energy calibration.

A quantitative comparison of the measurements to fixed-order NLO QCD calculations, corrected for non-perturbative and electroweak effects, shows overall fair agreement (with $p$-values in the percent range) when considering jet cross-sections in individual jet rapidity bins independently. In the inclusive jet measurement, a significant tension (with $p$-values $\ll10^{-3}$) between data and theory is observed when considering data points from all jet transverse momentum and rapidity regions. No significant differences between the inclusive jet cross-sections and the fixed-order NNLO QCD calculations corrected for non-perturbative and electroweak effects are observed when using $p_{\rm T}^{\mathrm{jet}}$ as the QCD scale. The NNLO pQCD predictions using $p_{\rm T}^{\mathrm{max}}$ as the scale overestimate the measured inclusive jet cross-sections.

\clearpage

\section*{Acknowledgements}


We thank CERN for the very successful operation of the LHC, as well as the
support staff from our institutions without whom ATLAS could not be
operated efficiently.

We acknowledge the support of ANPCyT, Argentina; YerPhI, Armenia; ARC, Australia; BMWFW and FWF, Austria; ANAS, Azerbaijan; SSTC, Belarus; CNPq and FAPESP, Brazil; NSERC, NRC and CFI, Canada; CERN; CONICYT, Chile; CAS, MOST and NSFC, China; COLCIENCIAS, Colombia; MSMT CR, MPO CR and VSC CR, Czech Republic; DNRF and DNSRC, Denmark; IN2P3-CNRS, CEA-DRF/IRFU, France; SRNSFG, Georgia; BMBF, HGF, and MPG, Germany; GSRT, Greece; RGC, Hong Kong SAR, China; ISF, I-CORE and Benoziyo Center, Israel; INFN, Italy; MEXT and JSPS, Japan; CNRST, Morocco; NWO, Netherlands; RCN, Norway; MNiSW and NCN, Poland; FCT, Portugal; MNE/IFA, Romania; MES of Russia and NRC KI, Russian Federation; JINR; MESTD, Serbia; MSSR, Slovakia; ARRS and MIZ\v{S}, Slovenia; DST/NRF, South Africa; MINECO, Spain; SRC and Wallenberg Foundation, Sweden; SERI, SNSF and Cantons of Bern and Geneva, Switzerland; MOST, Taiwan; TAEK, Turkey; STFC, United Kingdom; DOE and NSF, United States of America. In addition, individual groups and members have received support from BCKDF, the Canada Council, CANARIE, CRC, Compute Canada, FQRNT, and the Ontario Innovation Trust, Canada; EPLANET, ERC, ERDF, FP7, Horizon 2020 and Marie Sk{\l}odowska-Curie Actions, European Union; Investissements d'Avenir Labex and Idex, ANR, R{\'e}gion Auvergne and Fondation Partager le Savoir, France; DFG and AvH Foundation, Germany; Herakleitos, Thales and Aristeia programmes co-financed by EU-ESF and the Greek NSRF; BSF, GIF and Minerva, Israel; BRF, Norway; CERCA Programme Generalitat de Catalunya, Generalitat Valenciana, Spain; the Royal Society and Leverhulme Trust, United Kingdom.

The crucial computing support from all WLCG partners is acknowledged gratefully, in particular from CERN, the ATLAS Tier-1 facilities at TRIUMF (Canada), NDGF (Denmark, Norway, Sweden), CC-IN2P3 (France), KIT/GridKA (Germany), INFN-CNAF (Italy), NL-T1 (Netherlands), PIC (Spain), ASGC (Taiwan), RAL (UK) and BNL (USA), the Tier-2 facilities worldwide and large non-WLCG resource providers. Major contributors of computing resources are listed in Ref.~\cite{ATL-GEN-PUB-2016-002}.

\printbibliography
\clearpage



 
\begin{flushleft}
{\Large The ATLAS Collaboration}

\bigskip

M.~Aaboud$^\textrm{\scriptsize 34d}$,    
G.~Aad$^\textrm{\scriptsize 99}$,    
B.~Abbott$^\textrm{\scriptsize 124}$,    
O.~Abdinov$^\textrm{\scriptsize 13,*}$,    
B.~Abeloos$^\textrm{\scriptsize 128}$,    
S.H.~Abidi$^\textrm{\scriptsize 165}$,    
O.S.~AbouZeid$^\textrm{\scriptsize 143}$,    
N.L.~Abraham$^\textrm{\scriptsize 153}$,    
H.~Abramowicz$^\textrm{\scriptsize 159}$,    
H.~Abreu$^\textrm{\scriptsize 158}$
% ...
\bigskip
\\

% ...
$^{13}$Institute of Physics, Azerbaijan Academy of Sciences, Baku; Azerbaijan.\\
% ...
$^{34}$$^{(a)}$Facult\'e des Sciences Ain Chock, R\'eseau Universitaire de Physique des Hautes Energies - Universit\'e Hassan II, Casablanca;$^{(b)}$Centre National de l'Energie des Sciences Techniques Nucleaires (CNESTEN), Rabat;$^{(c)}$Facult\'e des Sciences Semlalia, Universit\'e Cadi Ayyad, LPHEA-Marrakech;$^{(d)}$Facult\'e des Sciences, Universit\'e Mohamed Premier and LPTPM, Oujda;$^{(e)}$Facult\'e des sciences, Universit\'e Mohammed V, Rabat; Morocco.\\
% ...
$^{99}$CPPM, Aix-Marseille Universit\'e, CNRS/IN2P3, Marseille; France.\\
% ...
$^{124}$Homer L. Dodge Department of Physics and Astronomy, University of Oklahoma, Norman OK; United States of America.\\
% ...
$^{128}$LAL, Universit\'e Paris-Sud, CNRS/IN2P3, Universit\'e Paris-Saclay, Orsay; France.\\
% ...
$^{143}$Santa Cruz Institute for Particle Physics, University of California Santa Cruz, Santa Cruz CA; United States of America.\\
% ...
$^{153}$Department of Physics and Astronomy, University of Sussex, Brighton; United Kingdom.\\
% ...
$^{158}$Department of Physics, Technion, Israel Institute of Technology, Haifa; Israel.\\
$^{159}$Raymond and Beverly Sackler School of Physics and Astronomy, Tel Aviv University, Tel Aviv; Israel.\\
% ...
$^{165}$Department of Physics, University of Toronto, Toronto ON; Canada.\\
% ...
$^{a}$ Also at  Department of Physics, University of Malaya, Kuala Lumpur; Malaysia.\\
$^{b}$ Also at Borough of Manhattan Community College, City University of New York, NY; United States of America.\\
$^{c}$ Also at Centre for High Performance Computing, CSIR Campus, Rosebank, Cape Town; South Africa.\\
$^{d}$ Also at CERN, Geneva; Switzerland.\\
$^{e}$ Also at CPPM, Aix-Marseille Universit\'e, CNRS/IN2P3, Marseille; France.\\
% ...
$^{*}$ Deceased

\end{flushleft}



\end{document}